\documentclass[aps,prd,12pt,preprint,superscriptaddress,nofootinbib,floatfix]{revtex4}
\usepackage{latexsym}
\usepackage{amsfonts}
\usepackage[latin1]{inputenc}
\usepackage[caption=false]{subfig}
\usepackage{graphicx}
\usepackage{mathrsfs}
\usepackage{amssymb}
\usepackage{amsmath}
\usepackage{verbatim}
\usepackage{multirow}
\usepackage{color}
\usepackage{epsfig}
\usepackage{amscd}
\usepackage{bm}
\usepackage{bbold}
\usepackage{axodraw}
\usepackage{slashed}
\usepackage{multirow}
\usepackage{booktabs}

\allowdisplaybreaks

\makeatletter
\renewcommand{\p@subsection}{}
\renewcommand{\p@subsubsection}{}
\makeatother

\newcommand{\nc}{\newcommand}

\nc{\newsection}[1]{\section{#1}\setcounter{equation}{0}}
\nc{\newappendix}[1]{\section{#1}\setcounter{equation}{0}}
\nc{\scm}{\scriptscriptstyle\mathrm}
\nc{\f}{\frac}
\nc{\be }{\begin{equation}}   \nc{\ee }{\end{equation}}
\nc{\bea}{\begin{eqnarray}}   \nc{\eea}{\end{eqnarray}}
\nc{\baa}{\begin{array}}      \nc{\eaa}{\end{array}}
\nc{\bit}{\begin{itemize}}    \nc{\eit}{\end{itemize}}
\nc{\ben}{\begin{enumerate}}  \nc{\een}{\end{enumerate}}
\nc{\bce}{\begin{center}}     \nc{\ece}{\end{center}}
\nc{\bfl}{\begin{flushright}} \nc{\efl}{\end{flushright}}
\nc{\btb}{\begin{tabular}}    \nc{\etb}{\end{tabular}}
\nc{\eps}{\varepsilon}
\nc{\vp}{\varphi}
\nc{\tvp}{\widetilde{\varphi}}
\nc{\D}{\mbox{$\not\!\!D$}}
\nc{\Db}{\mbox{${\raisebox{2mm}{\boldmath ${}^\leftarrow$}\hspace{-4mm} D}$}}
\nc{\Dfb}{\mbox{$\raisebox{2mm}{\boldmath ${}^\leftrightarrow$}\hspace{-4mm} D$}}
\nc{\vpj }{\mbox{${\vp^\dag i\,\raisebox{2mm}{\boldmath ${}^\leftrightarrow$}\hspace{-4mm} D_\mu\,\vp}$}}
\nc{\vpjt}{\mbox{${\vp^\dag i\,\raisebox{2mm}{\boldmath ${}^\leftrightarrow$}\hspace{-4mm} D_\mu^{\,I}\,\vp}$}}
\newcommand{\Eqn}[1]{Eq.~(\ref{#1})}
\newcommand{\Eqns}[2]{Eqs.~(\ref{#1}) and (\ref{#2})}

\newcommand{\GeV}{{\rm GeV}}

\begin{document}

\preprint{PSI-PR-14-07} 
\preprint{ZU-TH 25/14}
\date{\today}

\vspace*{1.0truecm}


\title{The $\mu\to e\gamma$ decay in a systematic effective \\
       field theory approach with dimension 6 operators}

\vspace*{1.0truecm}

\author{G.~M.~Pruna}\email[E-mail: ]{Giovanni-Marco.Pruna@psi.ch}
\affiliation{Paul Scherrer Institut, CH-5232 Villigen PSI, Switzerland}
\author{A.~Signer}\email[E-mail: ]{Adrian.Signer@psi.ch}
\affiliation{Paul Scherrer Institut, CH-5232 Villigen PSI, Switzerland}
\affiliation{Physik-Institut, Universit\"at Z\"urich, CH-8057 Z\"urich, Switzerland}

\begin{abstract}
\noindent
We implement a systematic effective field theory approach to the
benchmark process $\mu\to e \gamma$, performing automated one-loop
computations including dimension~6 operators and studying their
anomalous dimensions. We obtain limits on Wilson coefficients of a
relevant subset of lepton-flavour violating operators that contribute
to the branching ratio $\mu\to e \gamma$ at one-loop. In addition, we
illustrate a method to extract further constraints induced by the
mixing of operators under renormalisation-group evolution. This
results in limits on the corresponding Wilson coefficients directly at
the high scale. The procedure can be applied to other processes as
well and, as an example, we consider also lepton-flavour violating
decays of the $\tau$.
\end{abstract}

\maketitle

\newpage

\section{Introduction}
\label{sec:intro}
\setcounter{equation}{0} 
\noindent
The study of lepton-flavour violating (LFV) processes in the charged
sector offers a possibility to probe the Standard Model (SM) to very
high scales. Of particular importance is the LFV decay $\mu\to e
\gamma$. First, there are very impressive experimental limits on this
branching ratio. The current best limit ${\rm BR}(\mu^+\to e^+\gamma)
< 5.7\times 10^{-13}$~\cite{Adam:2013mnn} has been set by the MEG
collaboration at PSI and an upgrade of the experiment is underway to
improve the sensitivity further by an order of
magnitude~\cite{Baldini:2013ke}. Second, in the SM with neutrino
masses $m_\nu$ this branching ratio is suppressed by the tiny ratio
$(m_\nu/m_W)^4$, where $m_W$ is the mass of the $W$-boson.  Thus, the
SM branching ratio is well below any experimental limit that is
achievable in the foreseeable future and any positive signal for
$\mu\to e \gamma$ would be clear evidence for physics beyond the
Standard Model (BSM).  Conversely, improving limits on this branching
ratio would put even more serious constraints on many BSM
models. Given its importance the decay $\mu\to e \gamma$ has been
studied in a large number of explicit BSM models. Here, a more model
independent approach is taken.

The impact of a BSM model with new physics at a large energy scale
$\Lambda \gg m_W$ to observables at much smaller scales can be
described using an effective field theory (EFT) approach. The SM is
considered to be an EFT valid up to a scale $\Lambda$ and the BSM
effects at lower energies are described by operators of dimension
$n>4$, suppressed by powers of $\Lambda$. These operators are
generated from the BSM physics by integrating out the heavy non-SM
degrees of freedom. In general, the dominant effects are expected to
come from dimension~5 and dimension~6 operators. A minimal list of all
possible such operators formed from SM fields only and respecting the
$SU(3)\times SU(2)\times U(1)$ gauge invariance consists of one
dimension~5 operator~\cite{Weinberg:1979sa} and 64 dimension~6
operators~\cite{Buchmuller:1985jz, Grzadkowski:2010es}, five of which
are baryon number violating. As many of these operators actually
represent matrices in generation space, the total number of
coefficients needed to describe the most general case is rather
large. Nevertheless, this is a systematic approach to study the impact
of BSM physics to a large class of observables obtained from
experiments at very different energy scales. It is used in Higgs
physics, B-physics and the study of electric dipole moments to mention
just a few of the applications.

Applying these ideas to the flavour changing decay $\mu\to e \gamma$
we note that there is a dimension~6 operator ($Q_{e\gamma}$ to be
defined below) that induces such a decay directly at tree level. It is
clear that the MEG limit provides an extremely strong constraint on
the coefficient of this operator. However, such a decay can also be
induced indirectly from other operators that are not immediately
linked to $\mu\to e \gamma$. Thus, even if a particular BSM does not
induce the operator $Q_{e\gamma}$ at the high scale $\Lambda$, it can
lead to a non-vanishing contribution to $\mu\to e \gamma$. Broadly
speaking, this can happen in two different ways.

First, some dimension~6 operators other than $Q_{e\gamma}$ can induce
a decay $\mu\to e \gamma$ beyond tree level. The contribution to
$\mu\to e \gamma$ from dimension~6 operators at one loop has partially
been computed~\cite{Crivellin:2013hpa} and it has been found that
several operators contribute. This can lead to very serious
independent constraints on the coefficients of these operators.

The second possibility is through mixing in the renormalisation-group
(RG) evolution of the Wilson coefficient $C_{e\gamma}$ of the operator
$Q_{e\gamma}$. The Wilson coefficients $C_i(\Lambda)$ of the
higher-dimensional operators are determined at the high scale
$\Lambda$ by integrating out the heavy fields. If these coefficients
then are to be used to study the impact of the higher-dimensional
operators to observables at a lower scale $\lambda$, say $\lambda\sim
m_W$, the coefficients $C_i(m_W)$ have to be determined from
$C_i(\Lambda)$ through RG evolution. The one-loop RG evolution of the
dimension~6 operators has been studied~\cite{Jenkins:2013zja,
Jenkins:2013wua, Alonso:2013hga} and, as expected, it has been
found that other operators mix with $Q_{e\gamma}$ under the evolution.

The aim of this paper is to present a complete analysis of $\mu\to e
\gamma$ in the context of an EFT approach including dimension~6
operators. To this end, we repeat and extend the one-loop calculation
presented in~\cite{Crivellin:2013hpa} for this process with a RG
analysis. The RG running is done in two steps. We first evolve from
the large scale $\Lambda$ to the electroweak scale $m_V\sim m_W \sim
m_Z$ and then use a modified evolution suitable for the scales $m_\mu
\lesssim \lambda \lesssim m_Z$, where the mass of the muon, $m_\mu$,
is the scale at which the coefficient $C_{e\gamma}$ has to be
evaluated for the process $\mu\to e \gamma$.  We consider the subset
of all dimension~6 operators that are most directly linked to the LFV
decay. The details of the Lagrangian and the setup for the
calculations are given in Section~\ref{Sec:2}.  In Section~\ref{Sec:3}
the relation between the Lagrangian and the branching ratio is
discussed. Section~\ref{Sec:4} is the main part of the paper.
Section~\ref{Sec:4.1} starts with the one-loop result of the branching
ratio computed in the EFT.  The experimental limit on the branching
ratio can be translated directly into a limit for
$C_{e\gamma}(m_\mu)$.  From the explicit one-loop results, it is also
possible to extract limits on other Wilson coefficients evaluated at
the small scale. In a second step, in Section~\ref{Sec:4.2}, the
anomalous dimensions of the operator $Q_{e\gamma}$ and those operators
that mix with $Q_{e\gamma}$ are computed. These results are then used
to obtain limits on the Wilson coefficients of these operators,
evaluated directly at the large scale $\Lambda$. Our conclusions are
presented in Section~\ref{sec:conclusions}.  The details of the
renormalisation needed for the one-loop result and the anomalous
dimensions are given in Appendix~\ref{App:a}. In Appendix~\ref{App:b}
the result for the (unrenormalised) one-loop branching ratio is
listed. Finally, in Appendix~\ref{App:c} we apply the same method to
the LFV decays of the $\tau$ to obtain limits on the corresponding
Wilson coefficients.

\section{Effective D-6 extension of the SM: leptonic interactions}
\label{Sec:2}
\setcounter{equation}{0} 
\noindent
In this paper we take the point of view that the SM is an EFT valid up
to some large scale $\Lambda$ and BSM physics can be parametrised by
operators of dimension 6 (D-6). Higher dimensional operators are not
considered.  A complete list of gauge invariant D-6 operators has been
given, in~\cite{Grzadkowski:2010es}. In this section the subset of D-6
operators that are relevant for our analysis of $\mu\to e \gamma$ is
presented and the implementation of these operators in automated
computational tools is also briefly discussed.

\begin{table}[!ht] 
\centering
\renewcommand{\arraystretch}{1.5}
\begin{tabular}{||c|c||c|c||c|c||} 
\hline \hline
\multicolumn{2}{||c||}{$\psi^2 X\vp$} & 
\multicolumn{2}{|c||}{$\psi^2\vp^2 D$} &
\multicolumn{2}{|c||}{$\psi^2\vp^3$}\\
\hline
$Q_{eW}$  & $(\bar l_p \sigma^{\mu\nu} e_r) \tau^I \vp W_{\mu\nu}^I$&  
$Q_{\vp l}^{(1)}$  &$(\vpj)(\bar l_p \gamma^\mu l_r)$ &
$Q_{e\vp}$           & $(\vp^\dag \vp)(\bar l_p e_r \vp)$\\
$Q_{eB}$  & $(\bar l_p \sigma^{\mu\nu} e_r) \vp B_{\mu\nu}$&   
$Q_{\vp l}^{(3)}$ & $(\vpjt)(\bar l_p \tau^I \gamma^\mu l_r)$&
 & \\
 & &    
$Q_{\vp e}$  & $(\vpj)(\bar e_p \gamma^\mu e_r)$&
 & \\
\hline \hline
\end{tabular}
\caption{D-6 operators consisting of fermions and bosons, according to
  \cite{Grzadkowski:2010es}.\label{tab:no4ferm}}
\end{table}

\begin{table}[!ht]
\centering
\renewcommand{\arraystretch}{1.5}
\begin{tabular}{||c|c||c|c||c|c||c|c||}
\hline\hline
\multicolumn{2}{||c||}{$(\bar LL)(\bar LL)$} & 
\multicolumn{2}{|c||}{$(\bar RR)(\bar RR)$} &
\multicolumn{2}{|c||}{$(\bar LL)(\bar RR)$}&
\multicolumn{2}{|c||}{$(\bar LR)(\bar RL)$ and $(\bar LR)(\bar LR)$}\\
\hline
$Q_{ll}$    & $(\bar l_p \gamma_\mu l_r)(\bar l_s \gamma^\mu l_t)$ &
$Q_{ee}$    & $(\bar e_p \gamma_\mu e_r)(\bar e_s \gamma^\mu e_t)$ &
$Q_{le}$    & $(\bar l_p \gamma_\mu l_r)(\bar e_s \gamma^\mu e_t)$ &
$Q_{ledq}$   &  $(\bar l_p^j e_r)(\bar d_s q_t^j)$\\
$Q_{lq}^{(1)}$& $(\bar l_p \gamma_\mu l_r)(\bar q_s \gamma^\mu q_t)$&  
$Q_{eu}$     & $(\bar e_p \gamma_\mu e_r)(\bar u_s \gamma^\mu u_t)$&
$Q_{lu}$     & $(\bar l_p \gamma_\mu l_r)(\bar u_s \gamma^\mu u_t)$&
$Q_{lequ}^{(1)}$&$(\bar l_p^j e_r) \eps_{jk} (\bar q_s^k u_t)$ \\
$Q_{lq}^{(3)}$  & $(\bar l_p \gamma_\mu \tau^I l_r)(\bar q_s \gamma^\mu \tau^I q_t)$&
$Q_{ed}$     &$(\bar e_p \gamma_\mu e_r)(\bar d_s\gamma^\mu d_t)$ &
$Q_{ld}$     & $(\bar l_p \gamma_\mu l_r)(\bar d_s \gamma^\mu d_t)$&
$Q_{lequ}^{(3)}$&$(\bar l_p^j \sigma_{\mu\nu} e_r) \eps_{jk} (\bar q_s^k \sigma^{\mu\nu} u_t)$ \\
            &                                                  &
            &                                                  &
$Q_{qe}$     & $(\bar q_p \gamma_\mu q_r)(\bar e_s \gamma^\mu e_t)$&
            &                                                   \\
\hline\hline
\end{tabular}
\caption{D-6 operators consisting of four fermions, according to
  \cite{Grzadkowski:2010es}. \label{tab:4ferm}}
\end{table}

The Lagrangian considered in this paper is the SM Lagrangian
$\mathcal{L}_{\rm SM}$ extended by D-6 operators
\begin{align}
\mathcal{L} = \mathcal{L}_{\rm SM} + 
\frac{1}{\Lambda^2} \sum_i C_i\, Q_i ,
\label{Lag6}
\end{align}
where the sum is over the D-6 operators listed in
Tables~\ref{tab:no4ferm} and \ref{tab:4ferm}. These are the D-6
operators that can cause LFV interactions.  The dimension~5 operator
is not included in \Eqn{Lag6}: since the effect of this operator on
$\mu\to e\gamma$ transitions has been studied
before~\cite{Petcov:1976ff, Minkowski:1977sc} we do not consider it in
our analysis. The notation and conventions are taken
from~\cite{Grzadkowski:2010es}. In particular, $\{p,r,s,t\}$ denote
generation indices. In the Lagrangian the operators appear multiplied
by $C_i^{pr\ldots}/\Lambda^2$, where $C_i^{pr\ldots}$ are
dimensionless coefficient matrices with two or four generation
indices.  With regard to the Hermitian conjugation, it is worth to
remark that
\begin{itemize}
\item in the operator class $\psi^2\vp^2 D$, it is self-realised by
  transposition of generation indices;
\item in the operator classes $(\bar LL)(\bar LL)$, $(\bar RR)(\bar
  RR)$ and $(\bar LL)(\bar RR)$, it is self-realised by transposition
  of generation indices once the prescription $C^{prst}=C^{rpts}$ is
  assumed;
\item for the other operator classes, adding the Hermitian conjugate
  (not listed explicitly in
  Tables~\ref{tab:no4ferm}~and~\ref{tab:4ferm}) is understood.
\end{itemize}
Working in the physical basis rather than in the gauge basis, the two
operators of the $\psi^2 X\vp$ set are rewritten using
\begin{align}
Q_{eB}&\rightarrow Q_{e\gamma}c_W-Q_{eZ}s_W,\\
Q_{eW}&\rightarrow -Q_{e\gamma}s_W-Q_{eZ}c_W,
\end{align}
where $s_W=\sin(\theta_W)$ and $c_W=\cos(\theta_W)$ are the sine and
cosine of the weak mixing angle. 
The term
\begin{align}
\mathcal{L}_{e\gamma} \equiv 
\frac{C_{e\gamma}}{\Lambda^2} Q_{e\gamma} + \mbox{h.c.}
= \frac{C_{e\gamma}^{pr}}{\Lambda^2}
 (\bar l_p \sigma^{\mu\nu} e_r) \vp F_{\mu\nu} + \mbox{h.c.},
\end{align}
where $F_{\mu\nu}$ is the electromagnetic field-strength tensor, is
then the only term in the D-6 Lagrangian that induces a $\mu\to
e\gamma$ transition at tree level.  However, at one loop (and even
higher order) the other operators listed in Tables~\ref{tab:no4ferm}
and \ref{tab:4ferm} also potentially contribute. 

Finally, special attention is devoted to the operator $Q_{e\vp}$: in
Feynman gauge, the presence of such an operator produces Lagrangian
terms of the form
\begin{align}\label{eq:qef}
\mathcal{L}_{e\varphi}=& \frac{v^3}{2\sqrt{2}\Lambda^2}C_{e\varphi}^{pr}\bar{e}_pe_r+
\frac{3v^2}{2\sqrt{2}\Lambda^2}C_{e\varphi}^{pr}\bar{e}_pe_rh\nonumber \\
+& i\frac{v^2}{2\sqrt{2}\Lambda^2}C_{e\varphi}^{pr}\bar{e}_pe_r\widehat{Z}+
i\frac{v^2}{2\Lambda^2}C_{e\varphi}^{pr}\bar{e}_p\nu_r\widehat{W}^++\left[\dots\right].
\end{align} 
Apparently, this operator introduces Goldstone-boson
($\widehat{Z},\ \widehat{W}^\pm$) interactions which are not
compensated by any analogous vectorial term. However, the combination
of \Eqn{eq:qef} with the D-4 SM Yukawa terms gives
\begin{align}\label{eq:YukPlusEff}
\mathcal{L}_{\rm Yukawa}+\mathcal{L}_{e\varphi}=&
\frac{v}{\sqrt{2}}\left(
-y_{pr}+\frac{v^2}{2\Lambda^2}C_{e\varphi}^{pr}
\right)\bar{e}_pe_r\nonumber \\
+&\frac{1}{\sqrt{2}}\left(
-y_{pr}+\frac{v^2}{2\Lambda^2}C_{e\varphi}^{pr}
\right)\bar{e}_pe_rh+
\frac{v^2}{\sqrt{2}\Lambda^2}C_{e\varphi}^{pr}\bar{e}_pe_rh\nonumber \\
+&\frac{i}{\sqrt{2}}\left(
-y_{pr}+\frac{v^2}{2\Lambda^2}C_{e\varphi}^{pr}
\right)\bar{e}_pe_r\widehat{Z}
+ i\left(
-y_{pr}+\frac{v^2}{2\Lambda^2}C_{e\varphi}^{pr}
\right)\bar{e}_p\nu_r\widehat{W}^++\left[\dots\right].
\end{align}
From \Eqn{eq:YukPlusEff}, it is understood that any 3-point
off-diagonal interaction involving Goldstone bosons is not physical,
i.e. it can be removed by an orthogonal transformation. However, this
procedure results in
\begin{itemize}
\item a residual term with a physical Higgs supporting LFV currents;
\item a redefinition of the relation between leptonic Yukawa couplings
  and leptonic masses:
\begin{align}\label{YukRed}
y_{pp}\rightarrow 
\frac{\sqrt{2}m_p}{v}+\frac{v^2}{2\Lambda^2}C_{e\varphi}^{pp}.
\end{align}
\end{itemize}
In the framework of LFV processes at tree level and one loop, the
prescription of \Eqn{YukRed} is never relevant. However, it is of
fundamental importance in the case of flavour diagonal interactions
and related analyses such as the study of the anomalous magnetic
moment of the muon $(g-2)_\mu$.

In the following sections, one-loop calculations in the theory given
by the Lagrangian \Eqn{Lag6} will be presented. In order to perform
such calculations in an automated way, several openly available
tools were used:
\begin{itemize}
\item in order to obtain consistent Feynman rules, the described model
  was implemented both in {\tt LanHEP v3.1.9} \cite{Semenov:2010qt}
  and in {\tt FeynRules v2.0} \cite{Alloul:2013bka}, and the agreement
  among the two packages was checked;
\item in order to produce a model file for the {\tt FeynArts v3.9}
  \cite{Hahn:2000kx} and {\tt FormCalc v8.3}
  \cite{Hahn:1998yk,Nejad:2013ina} packages, the FeynArts interface of
  FeynRules was exploited;
\item the combined packages FeynArts/FormCalc were employed to
  generate non-integrated amplitudes to be elaborated afterwards with
  the symbolic manipulation system {\tt Form v4.0}
  \cite{Kuipers:2012rf}.
\end{itemize}

The list of resulting tree-level Feynman rules from the
Lagrangian~\Eqn{Lag6} is too long to be given explicitly in this
paper.  It will be provided after the publication of this work: it
will appear in the FeynRules model database\footnote{{\tt
http://feynrules.irmp.ucl.ac.be/wiki/ModelDatabaseMainPage}.} (in the
format of a FeynRules model file). However, the Feynman rule for the
$\mu-e-\gamma$ interaction (consisting of the effective tree-level
interaction plus the one-loop wave-function renormalisation (WFR) of
the relevant objects) is presented (see Appendix~\ref{App:a}).

\section{$\mu\to e\gamma$: Branching ratio and constraints}
\label{Sec:3}
\setcounter{equation}{0} 
\noindent
It is well known that in the limit $m_\mu\gg m_e$ the partial width
of the process $\mu\to e\gamma$ is given by
\begin{align}
\Gamma_{\mu\to e\gamma}=\frac{1}{16\pi m_\mu}\left|\mathcal{M}\right|^2,
\end{align}
where $\mathcal{M}$ is the transition amplitude, which contains the
model-dependent information. Computing $\mathcal{M}$ in the theory
given by \Eqn{Lag6} and confronting the corresponding branching ratio
${\rm BR}(\mu^+\to e^+\gamma)$ with the experimental
limit~\cite{Adam:2013mnn} allows to put constraints on the Wilson
coefficients $C_i$ of some of the D-6 operators in \Eqn{Lag6}.

To make this connection more explicit we note that the
Lagrangian \Eqn{Lag6} induces flavour-violating interactions $\mu\to
e\gamma$ that can be written as
\begin{align}
V^{\mu} =\frac{1}{\Lambda^2}
 i  \sigma^{\mu\nu} \left(C_{TL}\,\omega_L  
  + C_{TR}\,\omega_R\right)
\left(p_2\right)_\nu ,
\label{eq:Vform}
\end{align}
where the conventions described in Appendix~\ref{App:a} are used and
$\omega_{L/R}=1\mp\gamma^5$. Note that no term $\sim \gamma^\mu$
appears in \Eqn{eq:Vform} since such a term is forbidden by gauge
invariance. $C_{TL}$ and $C_{TR}$ are coefficients of dimension one
that depend on the Wilson coefficients of the D-6 operators and on the
parameters of the SM. The unpolarised squared matrix element is
expressed in terms of them as
\begin{align}
\left|\mathcal{M}\right|^2=
\frac{4 \left(|C_{TL}|^2+|C_{TR}|^2\right) m_\mu^4}{\Lambda^4},
\end{align}
and the branching ratio is
\begin{align}\label{BRanalitica}
{\rm BR}(\mu\to e\gamma)=\frac{\Gamma_{\mu\to e\gamma}}{\Gamma_\mu}=
\frac{m_\mu^3}{4\pi \Lambda^4\Gamma_\mu}\left(|C_{TL}|^2+|C_{TR}|^2\right)=
\frac{48 \pi ^2}{G_F^2 m_\mu^2}
\frac{\left(|C_{TL}|^2+|C_{TR}|^2\right)}{\Lambda^4 },
\end{align}
where $\Gamma_\mu=\left(G_F^2m_\mu^5\right)/\left(192\pi^3\right)$
is the SM total decay width of the muon. The result
\Eqn{BRanalitica} is well known in the literature, see
e.g. \cite{Marciano:2008zz} and references therein. Confronting this
result with the experimental upper limit~\cite{Adam:2013mnn}
established by the MEG collaboration on the $\mu^+\to e^+\gamma$
transition
\begin{align}\label{MEGlimit}
{\rm BR}(\mu^+\to e^+\gamma)\leq 5.7\cdot10^{-13},
\end{align}
the limit 
\begin{align}\label{EFFlimit_temp}
\frac{\sqrt{|C_{TL}|^2+|C_{TR}|^2}}{\Lambda^2}
\leq 4.3\cdot 10^{-14}\left[\GeV\right]^{-1}
\end{align}
can be obtained. 

At tree level, for the process $\mu^+\to e^+\gamma$ the coefficients
appearing in \Eqn{EFFlimit_temp} are given by $C_{TR}^{(0)} = - v\,
C^{\mu e}_{e\gamma}/\sqrt{2}$ and $C_{TL}^{(0)} = - v\,
(C^{e\mu}_{e\gamma})^*/\sqrt{2}$. In what follows, we will instead
compute the coefficients for the process $\mu^-\to e^-\gamma$ where
the tree-level results are given by $C_{TR}^{(0)} = - v\,
C^{e\mu}_{e\gamma}/\sqrt{2}$ and $C_{TL}^{(0)} = - v\, (C^{\mu
  e}_{e\gamma})^*/\sqrt{2}$. From now on the generation indices will
often be dropped and the simplified notation $C_{e\gamma}$ will be
used for either $C_{e\gamma}^{\mu e}$ or $C_{e\gamma}^{e \mu}$.
Similar remarks apply to $C_{eZ}$ and
$C_{e\varphi}$\footnote{However, for the sake of
    completeness, generation indices are retained in the results
    provided in Appendices~\ref{App:a}~and~\ref{App:b}.}. Applying
the constraint \Eqn{EFFlimit_temp} then immediately results in a
constraint on $C_{e\gamma}$.

It is clear that if the BSM physics is such that the matching at the
scale $\Lambda$ produces a sizable coefficient $C_{e\gamma}(\Lambda)$
this will be the dominant effect for ${\rm BR}(\mu\to e\gamma)$. On the
other hand it is perfectly possible that the coefficient
$C_{e\gamma}(\Lambda)$ is zero or strongly suppressed compared to
Wilson coefficients of other D-6 operators. In this case effects of
operators that enter $C_{TL}$ and $C_{TR}$ only at one loop can be
important. 

The result of $C_{TL}$ (or $C_{TR}$) computed at one loop can 
schematically be written as 
\begin{align}\label{Coneloop}
C_{TL}^{(1)} =  - \sqrt{2} v\,  \left(
C_{e\gamma} \left(1 + e^2 c_{e\gamma}^{(1)} \right)
+ \sum_{i\neq e\gamma} e^2 c_i^{(1)}\, C_i\right) ,
\end{align}
where the electromagnetic coupling $e$ stands for a generic coupling
and the coefficients $c_{e\gamma}^{(1)}$ and $c_i^{(1)}$ depend on SM
parameters such as $m_Z, m_l$ etc. To compute the branching ratio at
one loop, apart from wave-function renormalisation also the vacuum
expectation value (VEV) $v$ has to be renormalised. Even after this
renormalisation, the coefficients $c_{e\gamma}^{(1)}$ and $c_i^{(1)}$
in general contain ultraviolet singularities. These singularities have
to be absorbed by a renormalisation of the coefficient
$C_{e\gamma}$. By choosing a particular scheme for this subtraction, a
precise definition of the Wilson coefficient is given. In what
follows, the $\overline{\rm MS}$ scheme is used.

In passing, it should be mentioned that for the coefficient
$c_{e\gamma}^{(1)}$ also infrared singularities have to be taken into
consideration.  However, the primary interest of considering one-loop
corrections is in the contribution of operators other than
$Q_{e\gamma}$ to $C_{TL}$ and $C_{TR}$. The corrections $ \sim e^2
c_{e\gamma}^{(1)} C_{e\gamma}$ only result in a small modification of
the limit on $C_{e\gamma}$. Hence these corrections will not be
considered in this paper.

The renormalised Wilson coefficients and, therefore, the coefficients
$C_{TL}$ and $C_{TR}$ are scale dependent quantities.  Hence,
\Eqn{EFFlimit_temp} should be interpreted as a phenomenological
constraint on the Wilson coefficients at the relevant energy
scale. While $\lambda\sim m_\mu$ is the typical energy scale probed by
the MEG experiment, the explicit results presented in the next section
will show, that for some of the operators the relevant scale is the
electroweak scale $\lambda\sim m_V$.  In any case, these scales are
much lower than $\Lambda$, the natural scale for the Wilson
coefficients after integrating out the heavy non-SM fields. To stress
this subtlety \Eqn{EFFlimit_temp} is rewritten as
\begin{align}\label{EFFlimit}
\left. 
\frac{\sqrt{|C_{TL}(\lambda)|^2+|C_{TR}(\lambda)|^2}}{\Lambda^2}
\right|_{\lambda\ll \Lambda}\leq 4.3\cdot 10^{-14}\left[\GeV\right]^{-1}.
\end{align}

In the next section, the explicit result for the coefficients $C_{TL}$
and $C_{TR}$ of \Eqn{eq:Vform} computed in the context of the
Lagrangian~\Eqn{Lag6} at the tree level and one-loop level is
given. Furthermore, various contributions coming from different
operators are separately shown. Afterwards, the RG running of the
Wilson coefficients is studied and \Eqn{EFFlimit} is applied to obtain
bounds on each relevant coefficient at the scale $\Lambda$. These
limits provide the most direct link between the low-energy observable
${\rm BR}(\mu\to e\gamma)$ and BSM scenarios within an EFT framework.

\section{Results}
\label{Sec:4}
\setcounter{equation}{0} 
\noindent
In this section, analytical results and phenomenological studies
concerning the impact of \Eqn{EFFlimit} on the Wilson coefficients of
D-6 operators in the Lagrangian~\Eqn{Lag6} are presented. The study is
split into two parts:
\begin{itemize}
\item[{\bf 1}:] The complete result for the decay $\mu \to e\gamma$
in the EFT up to the one-loop level is calculated. These results are
then used to obtain bounds on the Wilson coefficients of D-6 operators
at the fixed scale $\lambda=m_\mu$ or $\lambda=m_V$, applying the
experimental constraint on the branching ratio ${\rm BR}(\mu \to
e\gamma)$.
\item[{\bf 2}:] The mixing of a subset of D-6 operators with
$Q_{e\gamma}$ under RG evolution is computed. Translating the
experimental constraint on ${\rm BR}(\mu \to e\gamma)$ to a limit on
$C_{e\gamma}(m_Z)$, bounds on Wilson coefficients $C_i(\Lambda)$ of
operators $Q_i$ that mix with $Q_{e\gamma}$ are then obtained. The
dependence on $\Lambda$ of these bounds is discussed.
\end{itemize}

Due to the high level of automation, a certain number of cross checks
was strongly required. Unless specified otherwise, every result of
this paper was tested under the following aspects:
\begin{itemize}
\item with no exceptions, all the calculations were performed in a
general $R_\xi$-gauge and it was verified that any physical result
is independent of the gauge parameters $\xi_\gamma$,  $\xi_W$, $\xi_Z$
and $\xi_G$; 
\item intermediate expansions or truncations were never applied,
i.e. only the complete and final result was expanded, to verify both
the gauge invariance up to any order of $1/\Lambda^2$ and the
numerical consistency of expansions with respect to the full result;
\item if possible, some quantities were computed in different
ways (e.g. the anomalous dimension of the operators $Q_{e\gamma}$ and
$Q_{eZ}$ were computed both with an Higgs boson in the final state and
its VEV), further checking the complete agreement between(among) the
two(many) results;
\item if possible, any non-original outcome was compared with previous
literature: in particular, SM results against \cite{Denner:1991kt,
Bardin:1999ak}, fixed order calculations
against \cite{Crivellin:2013hpa}, anomalous dimensions of the SM
parameters against \cite{Machacek:1983tz, Machacek:1983fi,
Machacek:1984zw} and anomalous dimensions of D-6 operators 
against \cite{Jenkins:2013zja, Jenkins:2013wua,
  Alonso:2013hga}\footnote{We thank the authors of
  \cite{Crivellin:2013hpa,Jenkins:2013zja, Jenkins:2013wua,
  Alonso:2013hga} for help in clarifying any  
  source of disagreement by private communication.}.
\end{itemize}

In the following subsections, analytical results and phenomenological
constraints are given.

\subsection{Branching ratio: results and constraints}
\label{Sec:4.1}
\noindent
In this subsection, the explicit results of the one-loop calculations
for the coefficients $C_{TL}$ and $C_{TR}$, i.e. the coefficients
$c_i^{(1)}$ as defined in \Eqn{Coneloop} are given. We use diagonal
Yukawa matrices throughout.

First of all, it was verified that no term $\sim \gamma^\mu$ is
generated by the Lagrangian \Eqn{Lag6} for the LFV interaction
$V^\mu$, as dictated by gauge invariance. Then, the tree-level and
one-loop results were calculated using standard techniques as
described in Section~\ref{Sec:3}.  Subsequently, the outcome was
expanded around $m_l\ll m_V$, i.e. considering the leptonic masses to
be much smaller than the bosonic ones. In this limit, the contribution
from the operator $Q_{e\vp}$ to $C_{TL}$ reads
\begin{align}
C_{TL} &=
C_{e\vp}^{\mu e}\frac{m_W s_W}{48 \sqrt{2} m_H^2 \pi ^2} 
\left(4 m_e^2+4 m_\mu^2+3 m_e^2 
\log\left[\frac{m_e^2}{m_H^2}\right]
+3 m_\mu^2 \log\left[\frac{m_\mu^2}{m_H^2}\right]\right)  \nonumber \\
&+ C_{e\vp}^{e \mu} \frac{m_W s_W}{48 \sqrt{2} m_H^2 \pi ^2} \,
\left( - m_e m_\mu \right) + \ldots\, ,
 \label{Cephi-full} 
\end{align}
where the ellipses stand for contributions from other operators. Since
$m_e\ll m_\mu$ we can drop the term proportional to $C_{e\vp}^{e\mu}$.
Keeping the term $\sim  m_e^2 C_{e\vp}^{\mu e}$ in \Eqn{Cephi-full}
ensures that the result for $C_{TR}$ can be obtained by $(\mu
\longleftrightarrow e)$. 

Finally, the complete set of LO contributions of D-6
operators in \Eqn{Lag6} (up to one-loop in SM couplings) was obtained
(see Table~\ref{tab:res}). The full result without expansion around
$m_l\ll m_V$ is lengthy and not suitable for a phenomenological
analysis, but is given (truncated at the order $1/\Lambda^2$) in
Appendix~\ref{App:b} , including the complete information about the
generation indices for the $Q_{e\gamma}$, $Q_{eZ}$ and $Q_{e\varphi}$
operators.
\begin{table}[!ht] 
\centering
\renewcommand{\arraystretch}{1.2}
\btb{||c|c|c||} 
\hline 
\hline 
Operator & \multicolumn{2}{|c||}{$C_{TL}$ or 
$C_{TR} (\mu \longleftrightarrow e)$} \\
\hline 
$Q_{e\gamma}$ & \multicolumn{2}{|c||}{
$
\begin{aligned}
-C_{e\gamma}\frac{\sqrt{2}  m_W s_W}{e}
\end{aligned}
$
} \\[2ex]
$Q_{eZ}$ & \multicolumn{2}{|c||}{
$
\begin{aligned}
-C_{eZ}\frac{ e m_Z }{16 \sqrt{2} \pi ^2}
\left(3-6 c_W^2+4 c_W^2 \log\left[\frac{m_W^2}{m_Z^2}\right]
+(12 c_W^2-6) \log\left[\frac{m_Z^2}{\lambda^2}\right] \right)
\end{aligned}
$
} \\[2ex]
$Q_{\vp l}^{(1)}$ & \multicolumn{2}{|c||}{
$
\begin{aligned}
-C_{\vp l}^{(1)}\frac{e m_e \left(1+s_W^2\right) }{24 \pi ^2}
\end{aligned}
$
} \\[2ex]
$Q_{\vp l}^{(3)}$ & \multicolumn{2}{|c||}{
$
\begin{aligned}
C_{\vp l}^{(3)}\frac{e m_e \left(3-2 s_W^2\right)}{48 \pi ^2}
\end{aligned}
$
} \\[2ex]
$Q_{\vp e}$ & \multicolumn{2}{|c||}{
$
\begin{aligned}
C_{\vp e}\frac{e m_\mu \left(3-2 s_W^2\right) }{48 \pi ^2}
\end{aligned}
$
} \\[2ex]
$Q_{e\vp}$ & \multicolumn{2}{|c||}{
$
\begin{aligned}
C_{e\vp}\frac{m_W s_W}{48 \sqrt{2} m_H^2 \pi ^2} 
\left(4 m_e^2+4 m_\mu^2+3 m_e^2 
\log\left[\frac{m_e^2}{m_H^2}\right]
+3 m_\mu^2 \log\left[\frac{m_\mu^2}{m_H^2}\right]\right)
\end{aligned}
$
} \\[2ex]
$Q_{lequ}^{(3)}$ & \multicolumn{2}{|c||}{
$
\begin{aligned}
-\frac{e}{2\pi^2}
\sum_um_u\left(C_{lequ}^{(3)}\right)^{\mu e uu}
\log{\left[\frac{m_u^2}{\lambda^2}\right]}
\end{aligned}
$} \\[2ex]
\hline \hline
Operator & $C_{TL}$ & $C_{TR}$ \\
\hline 
$Q_{le}$ & $
\begin{aligned}
\frac{e}{16 \pi ^2}\left(m_e C_{le}^{\mu eee}+m_\mu C_{le}^{\mu\mu\mu e}
+ m_\tau C_{le}^{\mu \tau\tau e}\right)
\end{aligned}
$ & $
\begin{aligned}
\frac{e}{16 \pi ^2}(m_e C_{le}^{eee\mu}+m_\mu C_{le}^{e\mu\mu\mu}
+m_\tau C_{le}^{e\tau\tau\mu})
\end{aligned}
$ \\[1ex]
 \hline
\hline
\etb
\caption{Complete set of results (up to one-loop) for the LO
 contributions of the various D-6 operators to the $\mu\to e\gamma$
 decay. For $C_{TL}$  ($C_{TR}$) the generation indices $\mu e$ ($e
 \mu$) are understood. \label{tab:res}}
\end{table}

The one-loop calculation leads to several UV-divergent terms in
connection with three operators: $Q_{e\gamma}$, $Q_{eZ}$ and
$Q_{lequ}^{(3)}$. After $\overline{\rm MS}$ renormalisation the
remnants of these UV singularities are logarithms with an electroweak
scale, $\log(m_V^2/\lambda^2)$, in the term proportional to $C_{eZ}$
and logarithms with the various quark mass scales,
$\log(m_u^2/\lambda^2)$ in the coefficient proportional to $C_{\mu
euu}^{(3)}\equiv (C_{lequ}^{(3)})^{\mu euu}$. The one-loop corrections
proportional to $C_{e\gamma}$ (not shown) also contain scale-dependent
logarithms. Thus, as expected the coefficients $C_{TL}$ and $C_{TR}$
are scale dependent.

The impact on the phenomenology of the scale evolution from the large
scale $\Lambda$ to the electroweak scale is studied in
Section~\ref{Sec:4.2}.  Here the coefficients are evaluated at the
small scale $\lambda \ll \Lambda$, in particular, $\lambda=m_Z$ for
$C_{eZ}$.  Thus, the result of Table~\ref{tab:res} can be combined
directly with \Eqn{EFFlimit} to put a limit on a set of coefficients
coming from 7 operators (out of the ensemble of 19, see
Tables~\ref{tab:no4ferm}~and~\ref{tab:4ferm}). The other operators do
not contribute to the tree-level or one-loop fixed scale result.

Under the assumption that only one Wilson coefficient at a time is
non-vanishing, the numerical limits of Table~\ref{tab:nres} are
obtained. They are given for the Wilson coefficients with generation
indices $\mu e$. Since we consider the unpolarised decay, the
corresponding limits with the generation indices $e\mu$ are of course
the same.  The numerical values of the input parameters have been
taken from the Particle Data Group review~\cite{Beringer:1900zz}.
Note that no limit on $C^{(3)}_{lequ}$ is given since its contribution
vanishes if evaluated at the natural scale $\lambda=m_u$. It is of
course possible that an interplay among the various coefficients leads
to cancellations that invalidate the limits given in
Table~\ref{tab:nres}. A possibility to pin down more specific limits
concerns the study of the correlation among various experimental
bounds (e.g., ${\rm BR}(Z\to e\mu)$, ${\rm BR}(\mu\to 3e)$, etc.), but
this is outside the scope of this work. Similarly, the study of
specific underlying theories that can lead to such cancellations is
outside the strict EFT framework we are using.

\begin{table}[t] 
\centering
\renewcommand{\arraystretch}{1.2}
\btb{||c|c||c|c||} 
\hline 
\hline 
3-P Coefficient & At fixed scale  & 4-P Coefficient & 
At fixed scale  \\
\hline 
$C_{e\gamma}^{\mu e}$ &\
$2.5\cdot 10^{-16}\frac{\Lambda^2}{\left[\GeV\right]^2}$ & 
$C_{le}^{\mu eee}$ &
 $4.4\cdot 10^{-8}\frac{\Lambda^2}{\left[\GeV\right]^2}$ \\
$C_{eZ}^{\mu e}(m_Z)$ & $1.4\cdot10^{-13}\frac{\Lambda^2}{\left[\GeV\right]^2}$ & 
$C_{le}^{\mu \mu\mu e}$ &\
 $2.1\cdot 10^{-10}\frac{\Lambda^2}{\left[\GeV\right]^2}$ \\
$C_{\vp l}^{(1)}$ &
$2.6\cdot10^{-10}\frac{\Lambda^2}{\left[\GeV\right]^2}$ & 
$C_{le}^{\mu \tau\tau e}$ & $1.2\cdot 10^{-11}\frac{\Lambda^2}{\left[\GeV\right]^2}$ \\
$C_{\vp l}^{(3)}$ &
$2.5\cdot10^{-10}\frac{\Lambda^2}{\left[\GeV\right]^2}$ & 
 &  \\
$C_{\vp e}$ & $2.5\cdot10^{-10}\frac{\Lambda^2}{\left[\GeV\right]^2}$
&  & \\
$C_{e\vp}^{\mu e}$ &
 $2.8\cdot10^{-8}\frac{\Lambda^2}{\left[\GeV\right]^2}$
&  & \\[1ex]
 \hline 
\hline
\etb
\caption{Limits on the Wilson coefficients contributing to the
$\mu\to e\gamma$ transition up to the one-loop level. \label{tab:nres}} 
\end{table}

The results of Tables~\ref{tab:res}~and~\ref{tab:nres} were partially
shown in the work of Crivellin, Najjari and
Rosiek~\cite{Crivellin:2013hpa}; in addition to their results, here a
complete treatment of the operators $Q_{eZ}$ and $Q_{e\vp}$ is
shown. Regarding the latter, a comment is required: the coefficient
$C_{e\vp}$ is connected to a two-loop Barr-Zee effect
\cite{Barr:1990vd}, and it is well known \cite{Chang:1993kw,
  Blankenburg:2012ex, Harnik:2012pb, Crivellin:2013wna,
  Crivellin:2014cta} that such a two-loop contribution could be of the
same order or even larger than the one-loop term of
Table~\ref{tab:res}. Even though such feature could surely be
relevant, its analysis is not a purpose of this paper.

\subsection{Anomalous dimensions: results and constraints}
\label{Sec:4.2}
\noindent
In the previous section, limits on the Wilson coefficients $C_i(m_V)$
or $C_i(m_l)$ have been obtained by a strict one-loop calculation.
However, the most direct information on the underlying BSM theory can
be obtained by information on the Wilson coefficients at the matching
scale, $C_i(\Lambda)$. Thus, the anomalous dimensions of the D-6
operators that are relevant for the (tree-level) $\mu\to e\gamma$
transition have to be studied. 

The anomalous dimensions of D-6 operators have been calculated
in~\cite{Jenkins:2013zja, Jenkins:2013wua, Alonso:2013hga}. We have
repeated the computations of those that are relevant to our case and
extended the treatment to include the running of the coefficient
$C_{e\gamma}(\lambda)$ to scales $\lambda < m_V$.

By direct computation, one finds that the running of the
$C_{e\gamma}^{\mu e}$ coefficient for $\lambda > m_V$ is governed by
\begin{align}
&\quad
16\pi^2\frac{\partial C_{e\gamma}^{\mu e}}{\partial \log{\lambda}}
\nonumber \\
&= \left(e^2\left(\frac{47}{3}+\frac{1}{4 c_W^2}-\frac{9}{4 s_W^2}\right)
+ 2\, Y_e^2+\left(\frac{1}{2} + 2 c_W^2 \right)Y_\mu^2
+\sum_l Y_l^2+3\sum_q Y_q^2\right)C_{e\gamma}^{\mu e} \nonumber \\
&+\left(6e^2\left(\frac{c_W }{s_W}-\frac{s_W}{c_W}\right)
-2c_W s_W Y_\mu^2\right)C_{eZ}^{\mu e}+16 e
\sum_uY_uC_{\mu e uu}^{(3)},
\label{eq:Cey}
\end{align}
and the  related quantity $C_{e\gamma}^{e\mu}$ can be obtained by
interchanging the generation indices,
i.e. $Y_\mu \longleftrightarrow Y_e$ and $C_{\mu e uu}^{(3)}
\longleftrightarrow C_{e\mu uu}^{(3)} $. Retaining only the dominant
terms, \Eqn{eq:Cey} becomes
\begin{align}\label{CEG_RGE}
&\quad
16\pi^2\frac{\partial C_{e\gamma}^{\mu e}}{\partial\log{\lambda}}
\nonumber \\
&\simeq \left(\frac{47 e^2}{3}+\frac{e^2}{4 c_W^2}
-\frac{9 e^2}{4 s_W^2}+3 Y_t^2\right)C_{e\gamma}^{\mu e}
+6e^2\left(\frac{c_W }{s_W}-\frac{s_W}{c_W}\right)C_{eZ}^{\mu e}
+16 e \sum_{u}Y_uC^{(3)}_{\mu e uu}.
\end{align}
From \Eqn{CEG_RGE}, it follows that direct contributions to the
evolution of $C_{e\gamma}$ come from the operator $Q_{e\gamma}$
itself, plus the orthogonal operator $Q_{eZ}$ and the four-fermion
operator $Q^{(3)}_{lequ}$. Of course, the corresponding coefficients
are precisely the UV singularities that appear in the renormalisation
of $C_{e\gamma}$, discussed in Section~\ref{Sec:4.1}.

In the same way, a similar structure for the RG running of the
$C_{eZ}^{\mu e}$ coefficient is found:
\begin{align}
&\quad
16\pi^2\frac{\partial C_{eZ}^{\mu e}}{\partial \log{\lambda}}\nonumber \\
&=\left(e^2\left(-\frac{47 }{3}+\frac{151 }{12 c_W^2}
-\frac{11 }{12 s_W^2}\right)+ 2\, Y_e^2
+\left(\frac{1}{2}+2 s_W^2 \right)Y_\mu^2+\sum_l Y_l^2+3\sum_q Y_q^2\right)C_{eZ}^{\mu e}
\nonumber \\
&-\left(\frac{2e^2}{3}\left(\frac{2 c_W }{s_W}
+\frac{31 s_W}{c_W}\right)+2c_W s_W Y_\mu^2\right)C_{e\gamma}^{\mu e}
+2e\left(\frac{3 c_W}{s_W}-\frac{5 s_W}{c_W}\right) 
\sum_uY_uC_{\mu euu}^{(3)}  \nonumber \\
&\simeq -\frac{2e^2}{3}\left(\frac{2 c_W }{s_W}
+\frac{31 s_W}{c_W}\right)C_{e\gamma}^{\mu e}+
\left(-\frac{47 e^2}{3}+\frac{151 e^2}{12 c_W^2}
-\frac{11 e^2}{12 s_W^2}+3 Y_t^2\right)C_{eZ}^{\mu e} \nonumber \\
&+2e\left(\frac{3 c_W}{s_W}-\frac{5 s_W}{c_W}\right)
\sum_uY_uC^{(3)}_{\mu e uu}.\label{CEZ_RGE}
\end{align}

From \Eqns{CEG_RGE}{CEZ_RGE}, it is understood that there is an
interplay in the evolution of $C_{e\gamma}$ and $C_{eZ}$. Moreover
their running is directly connected to $C^{(3)}_{\mu e uu}$. Hence, if
the underlying theory produces non-vanishing matching coefficients
$C^{(3)}_{e\mu uu}(\Lambda)$ they will induce an non-vanishing
$C_{e\gamma}(m_V)$, even if $C_{e\gamma}(\Lambda)$ happens to vanish.
In fact, there are even further operators that contribute indirectly
to $C_{e\gamma}(m_V)$, namely those operators that mix with
$Q^{(3)}_{lequ}$ under RG evolution. To include these in the analysis,
the contribution of operators listed in Tables~\ref{tab:no4ferm} and
\ref{tab:4ferm} to the anomalous dimension of $Q^{(3)}_{lequ}$ and
$Q^{(1)}_{lequ}$ have been evaluated. The corresponding coefficients
run according to
\begin{align}\label{C3_RGE}
16\pi^2\frac{\partial C^{(3)}_{\mu e tt}}{\partial\log{\lambda}}&\simeq 
\frac{7 e Y_t}{3}C_{e\gamma}^{\mu e}+
\frac{eY_t}{2}\left(
\frac{3 c_W }{ s_W}-\frac{5 s_W }{3 c_W}
\right)C_{eZ}^{\mu e}+\nonumber \\
&+\left(
\frac{2 e^2}{9 c_W^2}-\frac{3 e^2}{s_W^2}+\frac{3 Y_t^2}{2}
+\frac{8 g_S^2}{3}\right)C^{(3)}_{\mu ett}+
\frac{e^2}{8}  \left(\frac{5}{c_W^2}
+\frac{3}{s_W^2}\right)C^{(1)}_{\mu ett},\\
16\pi^2\frac{\partial C^{(1)}_{\mu ett}}{\partial\log{\lambda}}&\simeq
\left(\frac{30 e^2}{c_W^2}+\frac{18 e^2}{s_W^2}\right)C^{(3)}_{\mu ett}+
\left(
-\frac{11 e^2}{3 c_W^2}+\frac{15 Y_t^2}{2}-8 g_S^2
\right)C^{(1)}_{\mu ett}.\label{C1_RGE}
\end{align}
Supposing that the coefficients $C_{\mu euu}^{(3)}$, $C_{\mu e
  cc}^{(3)}$ and $C_{\mu ett}^{(3)}$ are of the same order, any
sub-leading term can be dropped by retaining only the top-Yukawa and
gauge couplings in the above equations. Combining
\Eqns{CEG_RGE}{CEZ_RGE} with \Eqns{C3_RGE}{C1_RGE}, a relatively simple
system of ordinary differential equations (SoODE) can be built and
used to study the impact of the operators in Tables~\ref{tab:no4ferm}
and \ref{tab:4ferm} to $\mu\to e \gamma$.

It should be noted that our analysis is restricted to the operators
listed in Tables~\ref{tab:no4ferm}~and~\ref{tab:4ferm} even though
there are additional D-6 operators that also contribute directly or
indirectly to the running of $C_{e\gamma}$ and
$C_{eZ}$~\cite{Jenkins:2013zja, Jenkins:2013wua, Alonso:2013hga}. In
principle, a complete analysis including all D-6 operators should be
performed, extending the SoODE presented above. However, the case of
the operator $Q^{(1)}_{lequ}$ presented in this analysis is the most
relevant one and serves as an illustration on how to obtain limits on
a large class of Wilson coefficients of operators that are not
directly related to the process under consideration.

Now that the SoODE is established, we can obtain limits on the various
Wilson coefficients. The main idea is as follows: an effective theory
is defined through its Wilson coefficients at some large scale
$\Lambda$. We will consider the relevant coefficients one-by-one,
i.e. setting $C_i(\Lambda) \neq 0$ and all the other $C_j(\Lambda)=0;
\ j\neq i$. Then we let the system evolve from $\lambda=\Lambda$ to
the electroweak scale $\lambda=m_V$. At this scale, we confront
$C_{e\gamma}(\lambda=m_V)$ with the experimental limit according to
Table~\ref{tab:nres}. This will result in a constraint on
$C_i(\Lambda)$. The same procedure could of course also be carried out
using $C_{eZ}(m_V)$ rather than $C_{e\gamma}(m_V)$.  However, the
corresponding limits on the various $C_i(\Lambda)$ would always be
less stringent.

It should also be mentioned that a rigorous application of EFT ideas
requires to properly evolve the fixed order coefficient $C_{e\gamma}$
from the scale $\lambda = m_\mu$ to $\lambda = m_V$. Obviously, the RG
equations given above are only applicable for the scales $\lambda >
m_V$. At the electroweak scale, another matching of the theory to a
second EFT should be made by integrating out the heavy SM fields, i.e.
the fields of mass $\sim m_V$, very similar to what is done in the
context of $B$ decays (see e.g. \cite{Buchalla:1995vs}). The new EFT,
valid for scales $\lambda < m_V$ then consists of operators with only
(light)quark- and lepton fields as well as gluons and the photon.  The
anomalous dimensions of these operators then have to be computed in
order to determine the complete running of the Wilson coefficient
$C_{e\gamma}$ for scales $m_\mu < \lambda < m_V$. As the numerical
effects of this procedure are rather modest, a somewhat simplified
analysis is performed. As previously investigated
in~\cite{Czarnecki:2001vf}, for the running of $C_{e\gamma}(\lambda)$
below the electroweak scale only the QED contributions are taken into
account. The corresponding RG equation reads
\begin{align}\label{CEG_RGElow}
&\quad 
16\pi^2\frac{\partial C_{e\gamma}}{\partial\log{\lambda}}
\simeq e^2 \left(10 + \frac{4}{3}\sum_q e_q^2(\lambda) \right)
C_{e\gamma}, 
\end{align}
where the contribution of four-fermion operators has been omitted and
$e_q(\lambda)$ denotes the electric charge of the fermion fields that
are dynamical at the scale $\lambda$. Applying \Eqn{CEG_RGElow} to the
value of $C_{e\gamma}^{\mu e}(m_\mu)$ (and
$C_{e\gamma}^{e\mu}(m_\mu)$) given in Table~\ref{tab:nres} we obtain
the limit
\begin{align}\label{CeyatmZ}
\sqrt{\frac{
|C_{e\gamma}^{\mu e}(m_Z)|^2 +|C_{e\gamma}^{e\mu}(m_Z)|^2}{2}}
 < 1.8\cdot 10^{-16} \frac{\Lambda^2}{\left[\GeV\right]^2} . 
\end{align}
This is the limit that will be used to determine the constraints on
the remaining Wilson coefficients at the scale $\Lambda$.

\begin{figure}[t]
\includegraphics[width=0.7\textwidth]{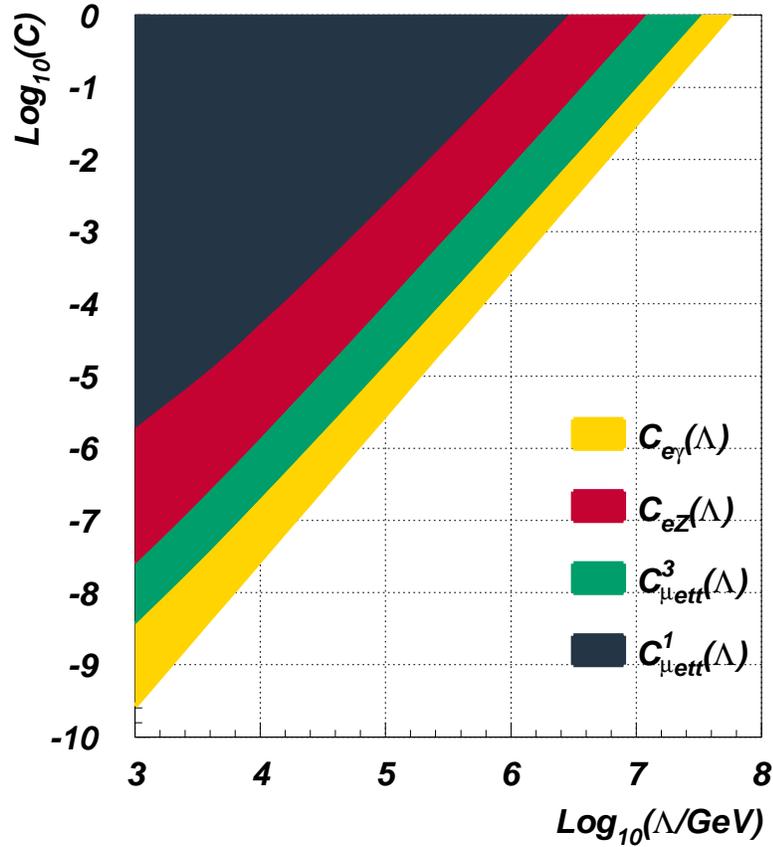}
\caption{Constraints on $C_{e\gamma}^{\mu e}$ (yellow), 
 $C_{eZ}^{\mu e}$ (green), $C^{(3)}_{\mu ett}$ (red) and $C^{(1)}_{\mu ett}$
(blue) plotted against the scale $\Lambda$ at which they are
defined. A $\log_{10}$-scale is adopted. The filled
area represents the excluded regions. }
\label{fig:RGE}
\end{figure}

In the RG evolution only the Yukawa coupling of the top is kept and
for all SM couplings one-loop running is implemented. Then the limits
on the Wilson coefficients $C_{e\gamma}$, $C_{eZ}$, $C^{(3)}_{\mu
ett}$ and $C^{(1)}_{\mu ett}$ are obtained as a function of the scale
$\Lambda$. The results are displayed in Figure~\ref{fig:RGE}. Not
surprisingly, the most severe constraint is on $C_{e\gamma}$ itself.
But also for $C_{eZ}$ and $C^{(3)}_{\mu ett}$ which affect the running
of $C_{e\gamma}$ directly, rather strong limits can be obtained. As
expected, the limits on $C^{(1)}_{\mu ett}$ are weaker, as it affects
$C_{e\gamma}$ only indirectly through $C^{(3)}_{\mu ett}$.

The dependence on $\Lambda$ of the limits on $C_{e\gamma}$ is close to
the canonical $\Lambda^2$ dependence, only slightly modified by the
running of the Wilson coefficients. For the other Wilson coefficients,
the effect of the running is somewhat larger. For illustrative
purposes, in Table~\ref{tab:rge_res}, the numerical values for the
Wilson coefficients for some choices of $\Lambda$ are given. Relaxing
the previous setup of only considering the top Yukawa coupling, the
analysis can also be extended to include $C_{\mu ecc}^{(3)}$ and
$C_{\mu ecc}^{(1)}$. Setting to zero all other Wilson coefficients at
$\Lambda$, in particular, $C_{\mu ett}^{(3)}(\Lambda)=0$ and $C_{\mu
ett}^{(1)}(\Lambda)=0$, it is then also possible to obtain limits on
$C_{\mu ecc}^{(3)}(\Lambda)$ and $C_{\mu ecc}^{(1)}(\Lambda)$. It is
clear that these limits get weaker with increasing $\Lambda$,
ultimately reaching the limit of perturbativity $\sim 4 \pi$. 

\begin{table}[t] 
\centering
\renewcommand{\arraystretch}{1.2}
\btb{||c||c|c|c||} 
\hline 
\hline 
3-P Coefficient & at  $\Lambda = 10^3~\GeV$ &
at  $\Lambda = 10^5~\GeV$ & at  $\Lambda = 10^7~\GeV$ \\
\hline 
$C_{e\gamma}^{\mu e}$ &  $2.7\cdot 10^{-10}$ & $2.9\cdot 10^{-6}$ & $3.1\cdot 10^{-2}$ \\
$C_{eZ}^{\mu e}$ & $2.5\cdot 10^{-8}$ & $1.0\cdot 10^{-4}$ &  $7.1\cdot 10^{-1}$  \\
$C^{(3)}_{\mu ett}$ & $3.6\cdot 10^{-9}$ & $1.4\cdot 10^{-5}$ & $9.8\cdot 10^{-2}$ \\
$C^{(1)}_{\mu ett}$ & $1.9\cdot 10^{-6}$ & $2.5\cdot 10^{-3}$ & n/a  \\
$C^{(3)}_{\mu ecc}$ & $4.8\cdot 10^{-7}$ & $1.9\cdot 10^{-3}$ & n/a \\
$C^{(1)}_{\mu ecc}$ & $2.6\cdot 10^{-4}$ & $3.3\cdot 10^{-1}$ & n/a \\
 \hline 
\hline
\etb
\caption{Limits on the Wilson coefficients defined at the scale
  $\lambda=\Lambda$ for three choices of $\Lambda =10^3, 10^5,
  10^7~\GeV$. \label{tab:rge_res} }
\end{table}

Besides this, other assumptions can be made less strict: while
\Eqns{CEG_RGE}{CEZ_RGE} are complete, sub-leading terms can be
gradually included in \Eqns{C3_RGE}{C1_RGE}. As an example,
reintroducing the bottom-Yukawa coupling and the CKM matrix
off-diagonal terms, the following leading contributions arise:
\begin{align}
16\pi^2\frac{\partial C^{(3)}_{\mu ett}}{\partial\log{\lambda}}&
\simeq \left[{\rm \Eqn{C3_RGE}}\right]
+Y_b^2V^\dagger_{33}\left(C^{(3)}_{\mu eut}V_{13}+C^{(3)}_{\mu ect}V_{23}\right)
+\left[\dots \right],\\
16\pi^2\frac{\partial C^{(1)}_{\mu ett}}{\partial\log{\lambda}}&
\simeq \left[{\rm \Eqn{C1_RGE}}\right]+2Y_bY_tC_{\mu ebb}
+\left[\dots \right],
\end{align}
where $C_{\mu ebb}$ is a coefficient related to the $Q_{ledq}$
operator, previously unconstrained.  However, as soon as one includes
other Yukawa couplings, the SoODE have to be enlarged to the point
that many other computations are required.  Nevertheless, in principle
the method can be systematised and generalised to including each
coefficient that could produce a (tree-level) $C_{e\gamma}$ transition
at the muonic mass scale, even if the contribution to the evolution is
not direct (as in the case of $C^{(1)}_{\mu ett}$).

To conclude this section, some limitations in our treatment are
mentioned (again). First, this analysis has been done in a strict
one-loop approximation, neglecting the possibility that for some
operators two-loop contributions could be more important. This can
happen in particular when through a two-loop effect a (small) Yukawa
coupling is replaced by gauge couplings, as is the case in the
Barr-Zee effect.

\begin{figure}[t]
\includegraphics[width=0.47\textwidth]{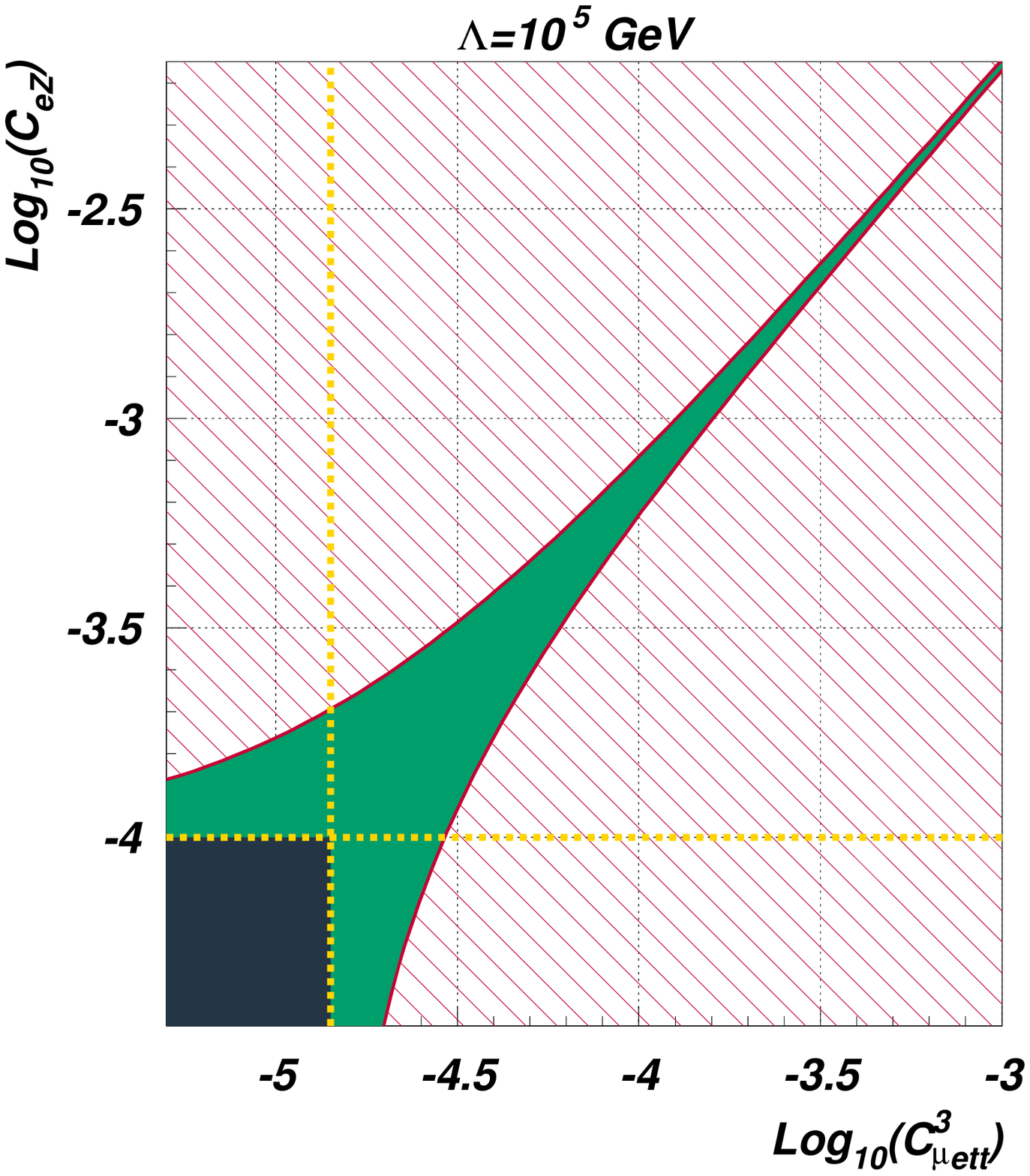}
\includegraphics[width=0.47\textwidth]{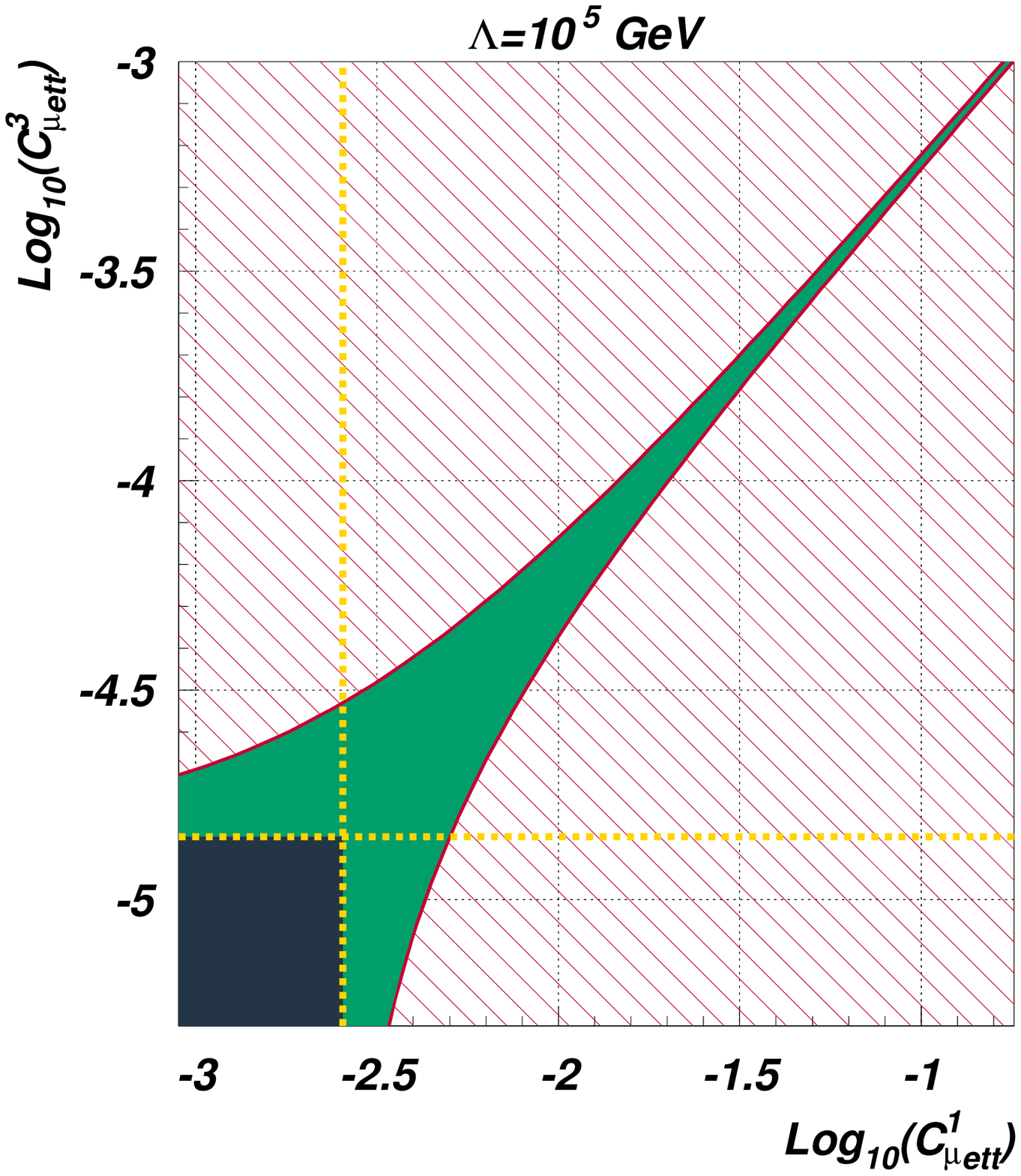}
\caption{Correlations between $C_{eZ}^{\mu e}$ and 
$C^{(3)}_{\mu ett}$ (left) and $C^{(3)}_{\mu ett}$ and $C^{(1)}_{\mu
ett}$ (right) at $\Lambda=10^5~\GeV$. The green area represents the
allowed regions if both coefficients are allowed to deviate from
zero.}
\label{fig:Corr}
\end{figure}

A second limitation regarding the limits presented in
Tables~\ref{tab:nres}~and~\ref{tab:rge_res} is that they have been
obtained assuming that only one coefficient at the time is
non-zero. It is clear that such an assumption is rather unrealistic. A
generic BSM model will usually introduce a large set of D-6 operators
when heavy fields are integrated out. Allowing for more than one
Wilson coefficient to be non zero, will introduce correlations that
can lead to allowed regions that clearly violate the limits given in
Tables~\ref{tab:nres}~and~\ref{tab:rge_res}. As an example we consider
the case when simultaneously $C_{eZ}(\Lambda)$ and $C^{(3)}_{\mu
  ett}(\Lambda)$ are non-vanishing (left panel of
Figure~\ref{fig:Corr}) as well as the case when simultaneously
$C^{(3)}_{\mu ett}(\Lambda)$ and $C^{(1)}_{\mu ett}(\Lambda)$ are
non-vanishing (right panel of Figure~\ref{fig:Corr}). The allowed
region (green) is clearly much larger than the allowed regions if only
one non-vanishing coupling at the time is allowed (indicated by the
yellow dotted lines). In principle, arbitrarily large values for
$C^{(3)}_{\mu ett}(\Lambda)$ are allowed, as long as $C_{eZ}(\Lambda)$
or $C^{(1)}_{\mu ett}(\Lambda)$ are tuned to provide an almost perfect
cancellation.  Such a fine-tuned choice of couplings is of course very
unnatural and at some point is in conflict with the fixed-order
constraint of $C_{eZ}$. Nevertheless, it has to be mentioned that the
limits presented in this analysis are to be taken more as guidelines
rather than strict limits. A more complete analysis with several
observables would be required to disentangle the correlations and get
strict limits on the various Wilson coefficients.

Finally, we recall that for $\lambda < m_V$ we have considered only
the running of $C_{e\gamma}$ induced by the pure QED contributions.
The effect of the running of $C_{e\gamma}$ from $\lambda=m_\mu$ to
$\lambda=m_V$ is below 10\% and we have checked that the impact of the
terms with Yukawa couplings is completely negligible. Hence, the use
of this approximation will affect the limits presented here by a few
percent at most. The only possible exception to this is the limit on
$C^{(3)}_{\mu ecc}$. As can be seen from \Eqn{eq:Cey}, if
$C^{(3)}_{\mu ecc}$ is much larger than $C_{e\gamma}$ the running of
$C_{e\gamma}$ for $m_c < \lambda < m_V$ is modified noticeably. Such a
situation can occur when considering the case $C^{(3)}_{\mu
  ecc}(\Lambda) \neq 0$ and all other $C_i(\Lambda)=0$, as done in
obtaining the limit on $C^{(3)}_{\mu ecc}$. In particular, if
$\Lambda$ is rather small, a very large $C^{(3)}_{\mu ecc}(\Lambda)$
is required to induce a sizable $C_{e\gamma}(m_V)$. We have checked
that, depending on the choice of $\Lambda$, the naive limits obtained
by having only $C^{(3)}_{\mu ecc}(\Lambda) \neq 0$ can be modified by
up to a factor two when taking into account its contribution to the RG
evolution for $\lambda < m_V$. The effect will be much smaller for a
more realistic scenario with several non-vanishing coefficients at the
large scale $\Lambda$.

\section{Conclusions}
\label{sec:conclusions}
\setcounter{equation}{0} 
\noindent
In this paper a complete one-loop analysis of the LFV decay $\mu\to e
\gamma$ in the context of an EFT with D-6 operators has been
presented. The main results are the limits on the (scale-dependent)
Wilson coefficients at the large matching scale. These limits provide
the most direct information on possible BSM models that can be
obtained from the $\mu\to e \gamma$ decay in an EFT framework.

It is not surprising that the limit on ${\rm BR}(\mu^+\to e^+ \gamma)$
results in a constraint on $C_{e\gamma}$, the Wilson
coefficient of the operator $Q_{e\gamma}$ that induces a tree-level $\mu\to e
\gamma$ transition.  What is more remarkable is that constraints can be
obtained also for a rather large number of further Wilson
coefficients. These belong to operators that indirectly induce a LFV
transition, either at one loop or through mixing under RG
evolution. In this context it is important to note that the Wilson
coefficients are scale dependent quantities and that in general
operators mix under RG evolution. Thus, the presence at the large
matching scale of any non-vanishing Wilson coefficient for an operator
that mixes with $Q_{e\gamma}$ under RG evolution will induce a LFV
transition $\mu\to e \gamma$ at the low scale. 

It is clear that such an analysis can be applied to other processes as
well. In particular, other LFV decays such as $\tau\to e\gamma$ or
$\tau\to \mu\gamma$ lead immediately to similar constraints for the
D-6 operators with other generation indices, as detailed in
Appendix~\ref{App:c}. But in principle, any observable for which there
are strong experimental constraints can be used. A combined analysis
with many observables will also potentially allow to disentangle
correlations between Wilson coefficients. Such correlations in the RG
running result in unnatural allowed regions which are governed by
large cancellations.

Depending on the process under consideration the inclusion of all D-6
operators, not only those listed in
Tables~\ref{tab:no4ferm}~and~\ref{tab:4ferm} might be required. While
this results in a more complicated system, such an analysis allows to
combine consistently experimental results that have been obtained at
completely different energy scales.  In the absence of clear evidence
for BSM physics at collider experiments, an extended EFT analysis
providing constraints on many Wilson coefficients directly at the
large scale can give useful clues in the search for a realistic BSM
scenario and we consider this to be a very promising and useful
strategy.

\section*{Acknowledgements} 
\setcounter{equation}{0} 
\noindent
The authors would like to thank A.~Crivellin and J.~Rosiek for most
helpful comments concerning the fixed order calculation and
theoretical details about the D-6 EFT. Furthermore, they gratefully
acknowledge R.~Alonso, E.~E.~Jenkins, A.~V.~Manohar and M.~Trott for
useful private communications with regards to the anomalous dimension
analysis of D-6 operators. GMP is thankful to C.~Duhr and C.~Degrande
for providing a constant and prompt help concerning the model file
implementation in the FeynRules package, as well as A.~Pukhov and
A.~Semenov for fruitful advices about the analogous task performed in
the framework of LanHEP. He is also grateful to T.~Hahn for detailed
support in the treatment of the four-fermion interactions in FormCalc
and to G.~W.~K\"alin for having extensively cross-checked the model
file.

The work of GMP has been supported by the European Community's Seventh
Framework Programme (FP7/2007-2013) under grant agreement n.~290605
(COFUND: PSI-FELLOW).

\clearpage

\appendix
\section{D-6 effective $\mu-e-\gamma$ interaction at one-loop: the Feynman rule}
\label{App:a}
\setcounter{equation}{0} 
\noindent
In this appendix, the Feynman rule for the $\mu-e-\gamma$ interaction
in the context of a D-6 ET is presented together with a complete
treatment of the LFV wave-function renormalisation.  Here and in
Appendix~\ref{App:b} we keep the generation indices of the Wilson
coefficients $C_{e\gamma}$, $C_{eZ}$ and $C_{e\vp}$, but for notational
simplicity drop the complex conjugate sign, i.e $(C_{e\gamma}^{\mu
  e})^* \to C_{e\gamma}^{\mu e}$.

In \Eqn{FRmegstruc}, the structure of the interaction is introduced in
terms of the new scale $\Lambda$ and four effective coefficients
related to the four possible contributions: vectorial left/right
($K_{VL}/K_{VR}$) and tensorial left/right ($K_{TL}/K_{TR}$). All
momenta are considered to be incoming.

\begin{center}
\fbox{
\begin{picture}(150,100)(0,0)
\SetWidth{1}
\SetColor{Black}
\ArrowLine(10,80)(75,50)
\Vertex(75,50){2}
\ArrowLine(75,50)(10,20)
\Photon(75,50)(140,50){5}{4}
\SetWidth{0.5}
\ArrowLine(140,60)(110,60)
\ArrowLine(30,20)(60,34)
\ArrowLine(30,80)(60,66)
\Text(10,90)[l]{$\mu(p_1)$}
\Text(10,10)[l]{$e(p_2-p_1)$}
\Text(140,70)[r]{$\gamma(-p_2)$}
\end{picture}
}
\begin{align}\label{FRmegstruc}
=\frac{1}{\Lambda^2}\left[\gamma^\mu\left(K_{VL}\, \omega_L + K_{VR}\, \omega_R\right)+\right.
\left.i  \sigma^{\mu\nu} \left(K_{TL}\, \omega_L  + K_{TR} \, \omega_R\right)
\left(p_2\right)_\nu\right].
\end{align}
\end{center}
The coefficients of \Eqn{FRmegstruc} are connected to the one-loop
wave-function renormalisation factors through
\begin{align}\label{FRmeg:1}
\frac{K_{VL}}{\Lambda^2}&=-\frac{e}{2}\left(\frac{1}{2}\delta Z_{e\mu}^L+\frac{1}{2}\left(\delta Z_{e\mu}^L\right)^\dagger\right)
-\frac{e v^2}{4 c_{W} s_{W}\Lambda^2}\left(C_{\varphi l}^{(1)}+C_{\varphi l}^{(3)}\right)\frac{1}{2}\delta Z_{ZA},\\
\label{FRmeg:2}
\frac{K_{VR}}{\Lambda^2}&=-\frac{e}{2}\left(\frac{1}{2}\delta Z_{e\mu}^R+\frac{1}{2}\left(\delta Z_{e\mu}^R\right)^\dagger\right)
-\frac{e v^2}{4 c_{W} s_{W}\Lambda^2}C_{\varphi e}\frac{1}{2}\delta Z_{ZA},\\
\label{FRmeg:3}
\frac{K_{TL}}{\Lambda^2}&=
-\frac{v}{\sqrt{2}\Lambda^2}C_{e\gamma}^{\mu e}
\left(1+\frac{1}{2}\delta Z_{\mu\mu}^L+\frac{1}{2}\left(\delta Z_{ee}^R\right)^\dagger
+\frac{1}{2}\delta Z_{AA}+\frac{\delta
  v}{v}\right)-
\frac{v}{\sqrt{2}\Lambda^2}C_{eZ}^{\mu e}\frac{1}{2}\delta Z_{ZA},\\
\label{FRmeg:4}
\frac{K_{TR}}{\Lambda^2}&=-\frac{v}{\sqrt{2}\Lambda^2}C_{e\gamma}^{e\mu}
\left(1+\frac{1}{2}\delta Z_{\mu\mu}^R+\frac{1}{2}\left(\delta Z_{ee}^L\right)^\dagger+\frac{1}{2}\delta Z_{AA}+\frac{\delta v}{v}\right)-\frac{v}{\sqrt{2}\Lambda^2}C_{eZ}^{e\mu}\frac{1}{2}\delta Z_{ZA}.
\end{align}
Several elements of Eqns.~(\ref{FRmeg:1})-(\ref{FRmeg:4}) do not
belong to the SM framework: the effective coefficients $C_{e\gamma}$,
$C_{eZ}$, $C_{\varphi l}^{(1)}$, $C_{\varphi l}^{(3)}$ and $C_{\varphi
e}$, plus the off-diagonal leptonic wave-function renormalisation. For
further information, a complete treatment of LFV wave-function
renormalisation in the on-shell scheme is given.

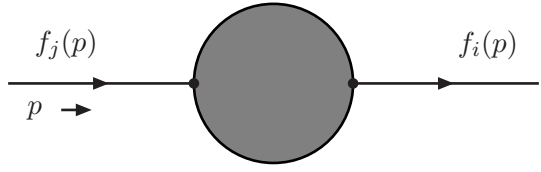
\begin{figure}[t]
\begin{center}
\begin{picture}(200,100)(0,0)
\SetWidth{1}
\SetColor{Black}
\GCirc(100,50){30}{0.5}
\SetColor{Black}
\ArrowLine(0,50)(70,50)
\LongArrow(20,40)(30,40)
\Text(10,40)[c]{$p$}
\Vertex(70,50){2}
\ArrowLine(130,50)(200,50)
\Vertex(130,50){2}
\Text(10,65)[l]{$f_j(p)$}
\Text(170,65)[l]{$f_i(p)$}
\end{picture}
\end{center}
\caption{Conventions used for the one-particle irreducible two-point
functions.}\label{ODLSE}  
\end{figure}

Making use of standard techniques (e.g., see \cite{Denner:1991kt}),
the off-diagonal leptonic self-energy (for conventions used see
Figure~\ref{ODLSE}) was calculated.  Then, the renormalisation
conditions in the on-shell scheme have been applied to obtain the
various contributions to the off-diagonal wave-function
renormalisation.
The tensorial structure that corresponds to such transition consists
of four possible coefficients: 
\begin{align}\label{SEstruc}
\Gamma^f_{ij}(p)=i\delta_{ij}(\slashed{p}-m_i)+i\left[
\slashed{p}\,\omega_L\,\Sigma^{f,L}_{ij}(p^2)+\slashed{p}\,\omega_R\,\Sigma^{f,R}_{ij}(p^2)+
\omega_L\,\Sigma^{f,l}_{ij}(p^2)+\omega_R\,\Sigma^{f,r}_{ij}(p^2)\right].
\end{align}
By applying the standard on-shell renormalisation conditions
\begin{align}
\left. {\rm Re}\!\left[\Gamma^{f}_{ij}(p)\right]u_j(p)\right|_{p^2=m_j^2}&=0,\\
\bar{u}_i(p){\rm Re}\!\left[\Gamma^{f}_{ij}(p)\right]_{p^2=m_i^2}&=0,
\end{align}
one finds the off-diagonal wave-function renormalisation that is
required in \Eqns{FRmeg:1}{FRmeg:2} to determine the
coefficients $K_{VL}$ and $K_{VR}$ of \Eqn{FRmegstruc}:
\begin{align}
\label{WFR:1}
\delta Z_{ij}^L&=\frac{4}{m^2_i-m^2_j}
\left(
m_j^2\Sigma^{f,L}_{ij}(m_j^2)+m_im_j\Sigma^{f,R}_{ij}(m_j^2)+
m_j\Sigma^{f,r}_{ij}(m_j^2)+m_i\Sigma^{f,l}_{ij}(m_j^2)
\right),\\
\label{WFR:2}
\delta Z_{ij}^R&=\frac{4}{m^2_i-m^2_j}
\left(
m_j^2\Sigma^{f,R}_{ij}(m_j^2)+m_im_j\Sigma^{f,L}_{ij}(m_j^2)+
m_j\Sigma^{f,l}_{ij}(m_j^2)+m_i\Sigma^{f,r}_{ij}(m_j^2)
\right).
\end{align}
The explicit result for the four coefficients of \Eqn{SEstruc} are as follows:
\begin{align}
&\quad \Sigma^{f,L}_{e\mu}(p^2)\Lambda^2\nonumber \\
&=A_0\left[m_e^2\right]\left(-\frac{m_e }{64 \pi ^2}C_{\varphi l}^{(1)}-\frac{m_e }{64 \pi ^2}C_{\varphi l}^{(3)}-\frac{3 m_Z  s_W^2}{16 \sqrt{2} \pi ^2}C_{eZ}^{\mu e}+\frac{3 m_Z  (2 s_W c_W)}{32 \sqrt{2} \pi ^2}C_{e\gamma}^{\mu e}\right) \nonumber \\
&+A_0\left[m_{\mu}^2\right]\left(\frac{m_\mu }{64 \pi ^2}C_{\varphi e}+\frac{3 m_Z  (c_W^2-s_W^2)}{32 \sqrt{2} \pi ^2}C_{eZ}^{\mu e}+\frac{3 m_Z  (2 s_W c_W)}{32 \sqrt{2} \pi ^2}C_{e\gamma}^{\mu e}\right) \nonumber \\
&+A_0\left[m_W^2\right]\left( \frac{3 m_Z  c_W^2}{16 \sqrt{2} \pi ^2}C_{eZ}^{\mu e}+\frac{3 m_Z  (2 s_W c_W)}{32 \sqrt{2} \pi ^2}C_{e\gamma}^{\mu e}\right) \nonumber \\
&+A_0\left[\xi_W m_W^2\right]\left( -\frac{m_e}{32 \pi ^2} C_{\varphi l}^{(3)}-\frac{m_Z  (2 s_W c_W)}{32 \sqrt{2} e \pi ^2}C_{e\varphi}^{\mu e}\right) \nonumber \\
&+A_0\left[m_Z^2\right]\left( \frac{3 m_Z  (-1+2 (c_W^2-s_W^2))}{32 \sqrt{2} \pi ^2}C_{eZ}^{\mu e}\right) \nonumber \\
&+A_0\left[\xi_Z m_Z^2\right]\left( \frac{m_\mu }{64 \pi ^2}C_{\varphi e}-\frac{m_e }{64 \pi ^2}C_{\varphi l}^{(1)}-\frac{m_e }{64 \pi ^2}C_{\varphi l}^{(3)}-\frac{m_Z  c_W s_W}{32 \sqrt{2} e \pi ^2}C_{e\varphi}^{\mu e}\right) \nonumber \\
&+A_0\left[m_H^2\right]\left(-\frac{3 m_Z  c_W s_W}{32 \sqrt{2} e \pi ^2}C_{e\varphi}^{\mu e}\right) \nonumber \\
&+B_0\left[p^2,m_Z^2,m_e^2\right]\left( \frac{3 m_e m_Z^2  s_W^2}{16 \pi ^2}C_{\varphi l}^{(1)}+\frac{3 m_e m_Z^2  s_W^2}{16 \pi ^2}C_{\varphi l}^{(3)}-\frac{3 m_Z \left(m_e^2+m_Z^2-p^2\right)  s_W^2}{16 \sqrt{2} \pi ^2}C_{eZ}^{\mu e}\right) \nonumber \\
&+B_0\left[p^2,\xi_Z m_Z^2,m_e^2\right]\left( \frac{m_e  \left(-m_e^2+\xi_Z m_Z^2+p^2-2 \xi_Z m_Z^2 (c_W^2-s_W^2)\right)}{64 \pi ^2}(C_{\varphi l}^{(1)}+C_{\varphi l}^{(3)})\right) \nonumber \\
&+B_0\left[p^2,m_Z^2,m_{\mu}^2\right]\left(-\frac{3 m_\mu m_Z^2  (c_W^2-s_W^2)}{32 \pi ^2}C_{\varphi e}+\frac{3 m_Z \left(m_\mu^2+m_Z^2-p^2\right)  (c_W^2-s_W^2)}{32 \sqrt{2} \pi ^2}C_{eZ}^{\mu e}\right) \nonumber \\
&+B_0\left[p^2,\xi_Z m_Z^2,m_{\mu}^2\right]\left( \frac{m_\mu  \left(m_\mu^2+\xi_Z m_Z^2-p^2-2 \xi_Z m_Z^2 (c_W^2-s_W^2)\right)}{64 \pi ^2}C_{\varphi e}\right) \nonumber \\
&+B_0\left[p^2,m_W^2,0\right]\left( \frac{3 m_Z \left(m_W^2-p^2\right)  c_W^2}{16 \sqrt{2} \pi ^2}C_{eZ}^{\mu e}+\frac{3 m_Z \left(m_W^2-p^2\right)  (2 s_W c_W)}{32 \sqrt{2} \pi ^2}C_{e\gamma}^{\mu e}\right) \nonumber \\
&+B_0\left[p^2,\xi_W m_W^2,0\right]\left( \frac{m_e \left(-\xi_W m_W^2+p^2\right) }{32 \pi ^2}C_{\varphi l}^{(3)}\right) \nonumber \\
&+B_0\left[p^2,m_H^2,m_e^2\right]\left( -\frac{m_e^2 m_Z  (2 s_W c_W)}{32 \sqrt{2} e \pi ^2}C_{e\varphi}^{\mu e}\right) \nonumber \\
&+B_0\left[p^2,m_H^2,m_{\mu}^2\right]\left(-\frac{m_\mu^2 m_Z  (2 s_W c_W)}{32 \sqrt{2} e \pi ^2}C_{e\varphi}^{\mu e}\right) \nonumber \\
&+B_0\left[p^2,0,m_e^2\right]\left( \frac{3 m_Z \left(m_e^2-p^2\right)  (2 s_W c_W)}{32 \sqrt{2} \pi ^2}C_{e\gamma}^{\mu e}\right) \nonumber \\
&+B_0\left[p^2,0,m_{\mu}^2\right]\left( \frac{3 m_Z \left(m_\mu^2-p^2\right) (2 s_W c_W)}{32 \sqrt{2} \pi ^2} C_{e\gamma}^{\mu e}\right)\nonumber \\
&+\frac{m_\mu m_Z^2  (c_W^2-s_W^2)}{16 \pi ^2}C_{\varphi e}-\frac{m_e m_Z^2  s_W^2}{8 \pi ^2}C_{\varphi l}^{(1)}-\frac{m_e m_Z^2  s_W^2}{8 \pi ^2}C_{\varphi l}^{(3)}\nonumber \\
&-\frac{m_Z}{16 \sqrt{2} \pi ^2}  \left(-2 \left(m_e^2-m_W^2+m_Z^2\right)+\left(2 m_e^2+2 m_\mu^2+2 m_W^2+4 m_Z^2-3 p^2\right) (c_W^2-s_W^2)\right)C_{eZ}^{\mu e}\nonumber \\
&-\frac{m_Z \left(2 m_e^2+2 m_\mu^2+2 m_W^2-3 p^2\right)  c_W s_W}{8 \sqrt{2} \pi ^2}C_{e\gamma}^{\mu e}.
\end{align}

\begin{align}
&\quad \Sigma^{f,R}_{e\mu}(p^2) =
\Sigma^{f,L}_{e\mu}(p^2) \big|_{\mu \leftrightarrow e}
\label{memswap}
\end{align}

\clearpage
\begin{align}
&\quad \Sigma^{f,l}_{e\mu}(p^2)\Lambda^2\nonumber \\
&=A_0\left[m_e^2\right]\left( -\frac{ \left(m_e^2+2 m_Z^2 (c_W^2-s_W^2)\right)}{64 p^2 \pi ^2}(C_{\varphi l}^{(1)}+C_{\varphi l}^{(3)})+\frac{m_e m_Z  c_W s_W}{32 \sqrt{2} e p^2 \pi ^2}C_{e\varphi}^{\mu e}\right.\nonumber \\
&+\left.\frac{3 m_e m_Z  (c_W^2-s_W^2)}{32 \sqrt{2} p^2 \pi ^2}C_{eZ}^{\mu e}+\frac{3 m_e m_Z (2 s_W c_W)}{32 \sqrt{2} p^2 \pi ^2}C_{e\gamma}^{\mu e} \right) \nonumber \\
&+A_0\left[m_{\mu}^2\right]\left(-\frac{ \left(m_\mu^2+2 m_Z^2 (c_W^2-s_W^2)\right)}{64 p^2 \pi ^2}(C_{\varphi l}^{(1)}+C_{\varphi l}^{(3)})+\frac{m_\mu m_Z  (2 s_W c_W)}{64 \sqrt{2} e p^2 \pi ^2}C_{e\varphi}^{e \mu}\right. \nonumber \\
&+\left.\frac{3 m_\mu m_Z  (c_W^2-s_W^2)}{32 \sqrt{2} p^2 \pi ^2}C_{eZ}^{e \mu}+\frac{3 m_\mu m_Z  (2 s_W c_W)}{32 \sqrt{2} p^2 \pi ^2}C_{e\gamma}^{e \mu}\right) \nonumber \\
&+A_0\left[m_W^2\right]\left(\frac{m_Z^2 \left(2 m_W^2-p^2\right)  c_W^2}{16 m_W^2 p^2 \pi ^2}C_{\varphi l}^{(3)}\right) \nonumber \\
&+A_0\left[\xi_W m_W^2\right]\left(\frac{m_Z^2  c_W^2}{16 m_W^2 \pi ^2}C_{\varphi l}^{(3)}\right) \nonumber \\
&+A_0\left[m_Z^2\right]\left( \frac{\left(m_e^2+m_\mu^2+4 m_Z^2-2 p^2\right)  (c_W^2-s_W^2)}{64 p^2 \pi ^2}(C_{\varphi l}^{(1)}+C_{\varphi l}^{(3)})\right.\nonumber \\
&-\left.\frac{3 m_\mu m_Z  (c_W^2-s_W^2)}{32 \sqrt{2} p^2 \pi ^2}C_{eZ}^{e \mu}-\frac{3 m_e m_Z  (c_W^2-s_W^2)}{32 \sqrt{2} p^2 \pi ^2}C_{eZ}^{\mu e}\right) \nonumber \\
&+A_0\left[\xi_Z m_Z^2\right] \left( \frac{ \left(m_e^2+m_\mu^2-\left(m_e^2+m_\mu^2-2 p^2\right) (c_W^2-s_W^2)\right)}{64 p^2 \pi ^2}(C_{\varphi l}^{(1)}+C_{\varphi l}^{(3)})\right) \nonumber \\
&+A_0\left[m_H^2\right]\left(-\frac{m_\mu m_Z  c_W s_W}{32 \sqrt{2} e p^2 \pi ^2}C_{e\varphi}^{e \mu}-\frac{m_e m_Z  c_W s_W}{32 \sqrt{2} e p^2 \pi ^2}C_{e\varphi}^{\mu e}\right) \nonumber \\
&+B_0\left[p^2,m_Z^2,m_e^2\right]\left( \frac{\left(m_e^4-2 m_Z^4+m_e^2 \left(m_Z^2-2 p^2\right)+m_Z^2 p^2+p^4\right)  (c_W^2-s_W^2)}{64 p^2 \pi ^2}(C_{\varphi l}^{(1)}+C_{\varphi l}^{(3)})\right.\nonumber\\
&+\left.\frac{3 m_e m_Z \left(-m_e^2+m_Z^2+p^2\right)  (c_W^2-s_W^2)}{32 \sqrt{2} p^2 \pi ^2}C_{eZ}^{\mu e}\right) \nonumber \\
&+B_0\left[p^2,\xi_Z m_Z^2,m_e^2\right]\left( \frac{1}{64 p^2 \pi ^2} \left(m_e^2 \left(m_e^2-\xi_Z m_Z^2-p^2\right)+\left(-m_e^4+\left(\xi_Z m_Z^2-p^2\right) p^2\right.\right.\right.\nonumber \\
&+\left.\left.\left.m_e^2 \left(\xi_Z m_Z^2+2 p^2\right)\right) (c_W^2-s_W^2)\right)(C_{\varphi l}^{(1)}+C_{\varphi l}^{(3)})\right) \nonumber \\
&+B_0\left[p^2,m_Z^2,m_{\mu}^2\right]\left(\frac{\left(m_\mu^4-2 m_Z^4+m_\mu^2 \left(m_Z^2-2 p^2\right)+m_Z^2 p^2+p^4\right)  (c_W^2-s_W^2)}{64 p^2 \pi ^2}(C_{\varphi l}^{(1)}+C_{\varphi l}^{(3)})\right.\nonumber \\ 
&+\left.\frac{3 m_\mu m_Z \left(-m_\mu^2+m_Z^2+p^2\right)  (c_W^2-s_W^2)}{32 \sqrt{2} p^2 \pi ^2}C_{eZ}^{e \mu}\right) \nonumber \\
&+B_0\left[p^2,\xi_Z m_Z^2,m_{\mu}^2\right]\left( \frac{1}{64 p^2 \pi ^2} \left(m_\mu^2 \left(m_\mu^2-\xi_Z m_Z^2-p^2\right)+\left(-m_\mu^4+\left(\xi_Z m_Z^2-p^2\right) p^2\right.\right.\right.\nonumber \\
&+\left.\left.\left. m_\mu^2 \left(\xi_Z m_Z^2+2 p^2\right)\right) (c_W^2-s_W^2)\right)(C_{\varphi l}^{(1)}+C_{\varphi l}^{(3)})\right) \nonumber \\
&+B_0\left[p^2,m_W^2,0\right]\left( -\frac{m_Z^2 \left(2 m_W^4-m_W^2 p^2-p^4\right)  c_W^2}{16 m_W^2 p^2 \pi ^2}C_{\varphi l}^{(3)}\right) \nonumber \\
&+B_0\left[p^2,\xi_W m_W^2,0\right]\left( \frac{m_Z^2 \left(\xi_W m_W^2-p^2\right)  c_W^2}{16 m_W^2 \pi ^2}C_{\varphi l}^{(3)}\right) \nonumber \\
&+B_0\left[p^2,m_H^2,m_e^2\right]\left( -\frac{m_e m_Z \left(m_e^2-m_H^2+p^2\right)  (2 s_W c_W)}{64 \sqrt{2} e p^2 \pi ^2}C_{e\varphi}^{\mu e}\right) \nonumber \\
&+B_0\left[p^2,m_H^2,m_{\mu}^2\right]\left( -\frac{m_\mu m_Z \left(-m_H^2+m_\mu^2+p^2\right) (2 s_W c_W)}{64 \sqrt{2} e p^2 \pi ^2}C_{e\varphi}^{e \mu} \right) \nonumber \\
&+ B_0\left[p^2,0,m_e^2\right]\left( -\frac{3 m_e m_Z \left(m_e^2-p^2\right)  (2 s_W c_W)}{32 \sqrt{2} p^2 \pi ^2}C_{e\gamma}^{\mu e}\right) \nonumber \\
&+ B_0\left[p^2,0,m_{\mu}^2\right]\left( -\frac{3 m_\mu m_Z \left(m_\mu^2-p^2\right)  (2 s_W c_W)}{32 \sqrt{2} p^2 \pi ^2}C_{e\gamma}^{e \mu}\right)\nonumber \\
&-\frac{m_Z^2  (c_W^2-s_W^2)}{16 \pi ^2}C_{\varphi l}^{(1)}-\frac{m_Z^2  (1+2 (c_W^2-s_W^2))}{16 \pi ^2}C_{\varphi l}^{(3)}-\frac{m_\mu m_Z  (c_W^2-s_W^2)}{16 \sqrt{2} \pi ^2}C_{eZ}^{e \mu}\nonumber \\
&-\frac{m_e m_Z  (c_W^2-s_W^2)}{16 \sqrt{2} \pi ^2}C_{eZ}^{\mu e}-\frac{m_\mu m_Z  c_W s_W}{8 \sqrt{2} \pi ^2}C_{e\gamma}^{e \mu}-\frac{m_e m_Z  c_W s_W}{8 \sqrt{2} \pi ^2}C_{e\gamma}^{\mu e}.
\end{align}

\begin{align}
&\quad \Sigma^{f,r}_{e\mu}(p^2)\Lambda^2\nonumber \\
&=A_0\left[m_e^2\right]\left(\frac{ \left(m_e^2+2 m_Z^2-2 m_Z^2 (c_W^2-s_W^2)\right)}{64 p^2 \pi ^2}C_{\varphi e}+\frac{m_e m_Z  c_W s_W}{32 \sqrt{2} e p^2 \pi ^2}C_{e\varphi}^{e \mu}\right.\nonumber \\
&-\left.\frac{3 m_e m_Z  s_W^2}{16 \sqrt{2} p^2 \pi ^2}C_{eZ}^{e \mu}+\frac{3 m_e m_Z  (2 s_W c_W)}{32 \sqrt{2} p^2 \pi ^2}C_{e\gamma}^{e \mu}\right) \nonumber \\
&+A_0\left[m_{\mu}^2\right]\left(\frac{ \left(m_\mu^2+2 m_Z^2-2 m_Z^2 (c_W^2-s_W^2)\right)}{64 p^2 \pi ^2}C_{\varphi e}+\frac{m_\mu m_Z  (2 s_W c_W)}{64 \sqrt{2} e p^2 \pi ^2}C_{e\varphi}^{\mu e}\right.\nonumber \\
&-\left.\frac{3 m_\mu m_Z  s_W^2}{16 \sqrt{2} p^2 \pi ^2}C_{eZ}^{\mu e}+\frac{3 m_\mu m_Z  (2 s_W c_W)}{32 \sqrt{2} p^2 \pi ^2}C_{e\gamma}^{\mu e}\right) \nonumber \\
&+A_0\left[m_Z^2\right]\left(-\frac{\left(m_e^2+m_\mu^2+4 m_Z^2-2 p^2\right)  s_W^2}{32 p^2 \pi ^2}C_{\varphi e}+\frac{3 m_e m_Z  s_W^2}{16 \sqrt{2} p^2 \pi ^2}C_{eZ}^{e \mu}+\frac{3 m_\mu m_Z  s_W^2}{16 \sqrt{2} p^2 \pi ^2}C_{eZ}^{\mu e}\right) \nonumber \\
&+A_0\left[\xi_Z m_Z^2\right]\left(-\frac{ \left(2 p^2+\left(m_e^2+m_\mu^2-2 p^2\right) (c_W^2-s_W^2)\right)}{64 p^2 \pi ^2}C_{\varphi e}\right) \nonumber \\
&+A_0\left[m_H^2\right]\left(-\frac{m_e m_Z  c_W s_W}{32 \sqrt{2} e p^2 \pi ^2}C_{e\varphi}^{e \mu}-\frac{m_\mu m_Z  c_W s_W}{32 \sqrt{2} e p^2 \pi ^2}C_{e\varphi}^{\mu e}\right) \nonumber \\
&+B_0\left[p^2,m_Z^2,m_e^2\right]\left(-\frac{\left(m_e^4-2 m_Z^4+m_e^2 \left(m_Z^2-2 p^2\right)+m_Z^2 p^2+p^4\right)  s_W^2}{32 p^2 \pi ^2}C_{\varphi e}\right.\nonumber\\
&-\left.\frac{3 m_e m_Z \left(-m_e^2+m_Z^2+p^2\right)  s_W^2}{16 \sqrt{2} p^2 \pi ^2}C_{eZ}^{e \mu}\right) \nonumber \\
&+B_0\left[p^2,\xi_Z m_Z^2,m_e^2\right]\left(\frac{1}{64 p^2 \pi ^2} \left(p^2 \left(-m_e^2-\xi_Z m_Z^2+p^2\right)+\left(-m_e^4+\left(\xi_Z m_Z^2-p^2\right) p^2\right.\right.\right.\nonumber \\
&+\left.\left.\left.m_e^2 \left(\xi_Z m_Z^2+2 p^2\right)\right) (c_W^2-s_W^2)\right)C_{\varphi e}\right) \nonumber \\
&+B_0\left[p^2,m_Z^2,m_{\mu}^2\right]\left(-\frac{\left(m_\mu^4-2 m_Z^4+m_\mu^2 \left(m_Z^2-2 p^2\right)+m_Z^2 p^2+p^4\right)  s_W^2}{32 p^2 \pi ^2}C_{\varphi e}\right.\nonumber \\
&-\left.\frac{3 m_\mu m_Z \left(-m_\mu^2+m_Z^2+p^2\right)  s_W^2}{16 \sqrt{2} p^2 \pi ^2}C_{eZ}^{\mu e}\right) \nonumber \\
&+B_0\left[p^2,\xi_Z m_Z^2,m_{\mu}^2\right]\left(\frac{1}{64 p^2 \pi ^2} \left(p^2 \left(-m_\mu^2-\xi_Z m_Z^2+p^2\right)+\left(-m_\mu^4+\left(\xi_Z m_Z^2-p^2\right) p^2\right.\right.\right.\nonumber\\
&+\left.\left.\left.m_\mu^2 \left(\xi_Z m_Z^2+2 p^2\right)\right) (c_W^2-s_W^2)\right)C_{\varphi e}\right) \nonumber \\
&+B_0\left[p^2,m_H^2,m_e^2\right]\left(-\frac{m_e m_Z \left(m_e^2-m_H^2+p^2\right)  (2 s_W c_W)}{64 \sqrt{2} e p^2 \pi ^2}C_{e\varphi}^{e \mu}\right) \nonumber \\
&+B_0\left[p^2,m_H^2,m_{\mu}^2\right]\left(-\frac{m_\mu m_Z \left(-m_H^2+m_\mu^2+p^2\right)  (2 s_W c_W)}{64 \sqrt{2} e p^2 \pi ^2}C_{e\varphi}^{\mu e}\right) \nonumber \\
&+B_0\left[p^2,0,m_e^2\right]\left(-\frac{3 m_e m_Z \left(m_e^2-p^2\right)  (2 s_W c_W)}{32 \sqrt{2} p^2 \pi ^2}C_{e\gamma}^{e \mu}\right) \nonumber \\
&+B_0\left[p^2,0,m_{\mu}^2\right]\left(-\frac{3 m_\mu m_Z \left(m_\mu^2-p^2\right)  (2 s_W c_W)}{32 \sqrt{2} p^2 \pi ^2}C_{e\gamma}^{\mu e}\right)\nonumber \\
&+\frac{m_Z^2  s_W^2}{8 \pi ^2}C_{\varphi e}-\frac{m_e m_Z  c_W s_W}{8 \sqrt{2} \pi ^2}C_{e\gamma}^{e \mu}-\frac{m_\mu m_Z  c_W s_W}{8 \sqrt{2} \pi ^2}C_{e\gamma}^{\mu e}\nonumber \\
&+\frac{m_e m_Z  s_W^2}{8 \sqrt{2} \pi ^2}C_{eZ}^{e \mu}+\frac{m_\mu m_Z  s_W^2}{8 \sqrt{2} \pi ^2}C_{eZ}^{\mu e}.
\end{align}

The explicit results for the four coefficients of the off-diagonal
one-particle irreducible two-point function for leptons are sufficient
to obtain the wave-function renormalisation factors
\Eqns{WFR:1}{WFR:2}.

Finally, for completeness we list the required SM expressions for the
renormalisation.  The expression
\begin{align}
\delta Z_{ZA}=\frac{\left(c_W e^2 \left(2 \left(-1+\xi_W^2\right)
m_W^2-(-9+\xi_W) A_0\left[m_W^2\right]-(5+3 \xi_W) A_0\left[\xi_W
m_W^2\right]\right)\right)}{\left(48 (-1+\xi_W) m_W^2 \pi ^2
s_W\right)}
\end{align}
is needed in Eqs.(\ref{FRmeg:1})-(\ref{FRmeg:4}) and the following expressions in
the $\overline{\rm MS}$ scheme are required for the computation of the
anomalous dimensions analysed in Section~\ref{Sec:4.2}:
\begin{align}
\widehat{\Delta}^{-1}\delta Z_{AA}&=-\frac{e^2 (20+3\xi_W)}{48 \pi ^2},\\
\widehat{\Delta}^{-1}\delta Z_{ZZ}&=-\frac{e^2 \left(-1+2 s_W^2+40 s_W^4+6 c_W^4\xi_W\right)}{96 \pi ^2 c_W^2s_W^2},\\
\widehat{\Delta}^{-1}\delta Z_{e}&=-\frac{1}{2}\delta Z_{AA}+\frac{s_W}{c_W}\frac{1}{2}\delta Z_{ZA}=\frac{11 e^2}{96 \pi ^2},\\
\widehat{\Delta}^{-1}\delta m_W^2&=+\frac{e^2 m_W^2}{32 \pi ^2 s_{W}^2}\xi_W+\frac{e^2 m_Z^2}{64 \pi ^2 s_{W}^2}\xi_Z \nonumber \\
&-\frac{e^2 \left(6\left(N_c \sum_qm_q^2+\sum_lm_l^2\right)+11 m_Z^2-20 m_Z^2 s_{W}^2\right)}{192 \pi ^2 s_{W}^2},\\
\widehat{\Delta}^{-1}\delta m_Z^2&=+\frac{e^2 m_W^2}{32 \pi ^2 c_W^2s_{W}^2}\xi_W+\frac{e^2 m_Z^2}{64 \pi ^2 c_W^2s_{W}^2}\xi_Z \nonumber \\
&-\frac{e^2 \left(6\left(N_c \sum_qm_q^2+\sum_lm_l^2\right)+11 m_Z^2-58 m_Z^2 s_{W}^2-44 m_Z^2 s_{W}^4\right)}{192 \pi ^2c_W^2 s_{W}^2},\\
\widehat{\Delta}^{-1}\frac{\delta v}{v}&=\left(1-\frac{1}{2 s_W^2}\right)\frac{\delta m_W^2}{m_W^2}+\frac{c_W^2 }{2 s_W^2}\frac{\delta m_Z^2}{m_Z^2}-\delta Z_{e}=\frac{e^2}{64 \pi ^2 s_W^2}\xi_W \nonumber \\
&+\frac{e^2 }{128 \pi ^2 c_W^2s_W^2}\xi_Z-\frac{e^2 \left(2\left(N_c \sum_qm_q^2+ \sum_lm_l^2\right)-9 m_Z^2+6 m_Z^2 s_W^2\right)}{128 m_Z^2 \pi ^2 c_W^2s_W^2 },\\
\widehat{\Delta}^{-1}\delta Z_{ll}^R&=-\frac{e^2 \left(m_l^2+2 m_Z^2 s_W^2 \left(c_W^2\xi_\gamma+ s_W^2\xi_Z\right)\right)}{32 m_Z^2 \pi ^2 c_W^2s_W^2},\\
\widehat{\Delta}^{-1}\delta Z_{ll}^L&=-\frac{e^2 \left(m_l^2+m_Z^2 \left(4 c_W^2s_W^2\xi_\gamma+2 c_W^2\xi_W+\left(1-2 s_W^2\right)^2\xi_Z \right)\right)}{64 m_Z^2 \pi ^2 c_W^2s_W^2 },\\
\widehat{\Delta}^{-1}\delta Z_{tt}^R&= -\frac{e^2 m_t^2 }{32 m_Z^2 \pi ^2 s_W^2 c_W^2}
-\frac{e^2 }{36 \pi ^2}\xi_\gamma
-\frac{e^2 s_W^2 }{36 \pi ^2 c_W^2}\xi_Z
-\frac{g_S^2}{12 \pi ^2} \xi_G,\\
\widehat{\Delta}^{-1}\delta Z_{tt}^L&= 
-\frac{e^2 \left(m_t^2+\sum_d \left|V_{d3}\right|^2 m_d^2 \right) }{64 m_Z^2 \pi ^2 s_W^2 c_W^2}
-\frac{e^2 }{36 \pi ^2}\xi_\gamma
-\frac{e^2 }{32 \pi ^2 s_W^2} \xi_W \nonumber \\
&-\frac{e^2 \left(3-4 s_W^2\right)^2 }{576 \pi ^2 s_W^2 c_W^2}\xi_Z
-\frac{g_S^2}{12 \pi ^2} \xi_G,\\
\widehat{\Delta}^{-1}\delta Z_{ct}^R&=
-\frac{e^2 m_c^2 m_t^2 \sum_dV_{3d} V_{2d}^\dagger}{32 m_Z^2 \pi ^2 s_W^2 c_W^2}=0,\\
\widehat{\Delta}^{-1}\delta Z_{ct}^L&=
-\frac{e^2 \sum_dV_{3d} V_{2d}^\dagger m_d^2 }{32 m_Z^2 \pi ^2 s_W^2 c_W^2}
-\frac{e^2  \sum_dV_{3d}V_{2d}^\dagger}{16 \pi ^2 s_W^2 }\xi_W=
-\frac{e^2 \sum_dV_{3d} V_{2d}^\dagger m_d^2 }{32 m_Z^2 \pi ^2 s_W^2 c_W^2},
\end{align}
where
\begin{align}
\widehat{\Delta}=\left[\frac{2}{4-D}-\gamma_E+\log{4\pi}\right],
\end{align}
with $D$ being the dimensional-regularisation parameter and $\gamma_E$
the Euler-Mascheroni constant. All the above equations have been cross
checked against \cite{Denner:1991kt}\footnote{In the Feynman Gauge,
i.e. $\xi \rightarrow 1$.} and \cite{Bardin:1999ak}.

\section{Explicit one-loop result for $\mu-e-\gamma$ }
\label{App:b}
\setcounter{equation}{0} 
\noindent
In this appendix, the complete result for the unrenormalised
coefficients $\bar{C}_{TL}$ and $\bar{C}_{TR}$ of the $\mu^-\to e^-
\gamma$ decay in the EFT is given.  After renormalisation, the
formulae were further expanded around $m_l\ll m_V$ to obtain the
results in Table~\ref{tab:res}; then the public package {\tt LoopTools
  2.10} \cite{Hahn:1998yk} was used to check the numerical stability
of the aforementioned expansion.  The result is presented in terms of
Passarino-Veltman functions \cite{Passarino:1978jh}, following the
convention described in \cite{Denner:1991kt}. Writing the coefficients as
\begin{align}
\bar{C}_{TL}&=C_{TL}^{(A_0)}+C_{TL}^{(B_0)}+C_{TL}^{(C_0)}+C_{TL}^{(c)},\\
\bar{C}_{TR}&=\bar{C}_{TL}\big|_{e \leftrightarrow \mu} , \label{swap}
\end{align}
the results read
\begin{align}
&\quad C_{TL}^{(A_0)}\nonumber \\
&=A_0\left[m_e^2\right]\left(
\frac{e \left(m_e^2-m_\mu^2+4 m_Z^2 s_W^2\right)}{64 m_\mu \left(m_e^2-m_\mu^2\right) \pi ^2}C_{\varphi e}+
\frac{e m_Z^2 \left(1-2 s_W^2\right)}{32 \left(m_e^3-m_e
m_\mu^2\right) \pi ^2}(C_{\varphi l}^{(1)}+C_{\varphi l}^{(3)}) \right. \nonumber\\
&-\left. \frac{ m_e m_W s_W}{32 \sqrt{2} \left(-m_e^2 m_\mu+m_\mu^3\right) \pi ^2}C_{e\varphi}^{e\mu}+
\frac{ m_\mu m_W s_W}{32 \sqrt{2} \left(-m_e^2 m_\mu+m_\mu^3\right) \pi ^2}C_{e\varphi}^{\mu e}\right. \nonumber \\
&+\left. \frac{ e m_\mu m_Z \left(3-4 s_W^2\right)}{32 \sqrt{2} \left(-m_e^2 m_\mu+m_\mu^3\right) \pi ^2}C_{eZ}^{\mu e}+
\frac{e m_Z \left(m_e+4 m_e s_W^2\right)}{32 \sqrt{2} \left(-m_e^2 m_\mu+m_\mu^3\right) \pi ^2}C_{eZ}^{e\mu}\right)\nonumber \\
&+A_0\left[m_\mu^2\right]\left(
\frac{e m_Z^2 s_W^2}{16 m_e^2 m_\mu \pi ^2-16 m_\mu^3 \pi ^2}C_{\varphi e}+
\frac{e \left(-m_e^2+m_\mu^2+2 m_Z^2 \left(1-2
s_W^2\right)\right)}{64 \left(m_e^3-m_e m_\mu^2\right) \pi
^2}(C_{\varphi l}^{(1)}+C_{\varphi l}^{(3)}) \right. \nonumber\\
&+\left. \frac{ m_e m_W s_W}{32 \sqrt{2} \left(m_e^3-m_e m_\mu^2\right) \pi ^2}C_{e\varphi}^{\mu e}
-\frac{ m_\mu m_W s_W}{32 \sqrt{2} \left(m_e^3-m_e m_\mu^2\right) \pi ^2}C_{e\varphi}^{e\mu}\right.\nonumber \\
&-\left. \frac{e m_\mu m_Z \left(3-4 s_W^2\right)}{32 \sqrt{2} \left(m_e^3-m_e m_\mu^2\right) \pi ^2}C_{eZ}^{e\mu}-\frac{ e m_Z \left(m_e+4 m_e s_W^2\right)}{32 \sqrt{2} \left(m_e^3-m_e m_\mu^2\right) \pi ^2}C_{eZ}^{\mu e}\right) \nonumber \\
&+A_0\left[m_W^2\right]\left(
\frac{e \left(m_e^2+2 m_Z^2 \left(-1+s_W^2\right)\right)}{16 \left(m_e^3-m_e m_\mu^2\right) \pi ^2}C_{\varphi l}^{(3)}\right. \nonumber \\
&+\left.\frac{e \left(-3 m_e^2+3 m_\mu^2-m_W^2\right)}{16 \sqrt{2} \left(-m_e^2+m_\mu^2\right) m_Z \pi ^2}C_{eZ}^{\mu e}-\frac{c_W^2 e m_e m_Z}{16 \sqrt{2} m_\mu \left(-m_e^2+m_\mu^2\right) \pi ^2}C_{eZ}^{e\mu}
\right) \nonumber \\
&+A_0\left[m_Z^2\right]\left(
\frac{e \left(m_e^2-m_\mu^2+8 m_Z^2 s_W^2\right)}{64 \left(-m_e^2 m_\mu+m_\mu^3\right) \pi ^2}C_{\varphi e}+
\frac{e \left(m_e^2-m_\mu^2+4 m_Z^2 \left(-1+2 s_W^2\right)\right)}{64 \left(m_e^3-m_e m_\mu^2\right) \pi ^2}(C_{\varphi l}^{(1)}+C_{\varphi l}^{(3)}) \right.\nonumber \\
&+\left. \frac{e m_Z}{8 \sqrt{2} \left(m_e^2-m_\mu^2\right) \pi ^2}C_{eZ}^{\mu e}
+\frac{e m_Z \left(m_\mu^2 \left(3-4 s_W^2\right)+m_e^2 \left(1+4 s_W^2\right)\right)}{32 \sqrt{2} m_e m_\mu \left(m_e^2-m_\mu^2\right) \pi ^2}C_{eZ}^{e\mu}
\right)\nonumber \\
&-A_0\left[m_H^2\right]\frac{m_W s_W}{32 \sqrt{2} m_e m_\mu \pi ^2}C_{e\varphi}^{e\mu},
\end{align}

\begin{align}
&\quad C_{TL}^{(B_0)}\nonumber \\
&=B_0\left[m_e^2,0,m_W^2\right]\left(-\frac{e \left(2 m_\mu^2 m_W^4+m_e^4 \left(m_\mu^2+4 m_W^2\right)-m_e^2 m_W^2 \left(5 m_\mu^2+4 m_W^2\right)\right) }{16 m_e \left(m_e^2-m_\mu^2\right)^2 \pi ^2}C_{\varphi l}^{(3)} \right.\nonumber \\
&-\left. \frac{c_W^2 e m_e m_\mu m_Z \left(m_e^2+m_Z^2-m_Z^2 s_W^2\right)}{16 \sqrt{2} \left(m_e^2-m_\mu^2\right)^2 \pi ^2}C_{eZ}^{e\mu}+\frac{c_W^2 e \left(4 m_e^4+3 m_\mu^2 m_W^2+m_e^2 \left(-5 m_\mu^2-4 m_W^2\right)\right) m_Z}{16 \sqrt{2} \left(m_e^2-m_\mu^2\right)^2 \pi ^2}C_{eZ}^{\mu e}\right) \nonumber \\
&+B_0\left[m_e^2,m_e^2,m_H^2\right]\left(-\frac{m_e \left(2 m_e^2-m_H^2\right) m_\mu m_W s_W}{32 \sqrt{2} \left(m_e^2-m_\mu^2\right)^2 \pi ^2}C_{e\varphi}^{e\mu}
+\frac{\left(2 m_e^2-m_H^2\right) \left(2 m_e^2-m_\mu^2\right) m_W s_W}{32 \sqrt{2} \left(m_e^2-m_\mu^2\right)^2 \pi ^2}C_{e\varphi}^{\mu e}
\right) \nonumber \\
&+B_0\left[m_e^2,m_e^2,m_Z^2\right]\left(
\frac{e m_\mu \left(-2 m_e^4+m_e^2 \left(2 m_\mu^2-m_Z^2\right)+m_Z^2 \left(4 m_Z^2 s_W^2+m_\mu^2 \left(1-8 s_W^2\right)\right)\right)}{64 \left(m_e^2-m_\mu^2\right)^2 \pi ^2}C_{\varphi e}\right.\nonumber \\
&-\left.\frac{e m_Z^2 \left(3 m_e^4+m_\mu^2 m_Z^2 \left(1-2 s_W^2\right)-m_e^2 \left(2 m_Z^2 \left(1-2 s_W^2\right)+m_\mu^2 \left(1+4 s_W^2\right)\right)\right)}{32 m_e \left(m_e^2-m_\mu^2\right)^2 \pi ^2}(C_{\varphi l}^{(1)}+C_{\varphi l}^{(3)}) \right.\nonumber \\
&+ \left.\frac{e m_e m_\mu m_Z \left(2 m_e^2-m_Z^2\right) \left(1+4 s_W^2\right)}{32 \sqrt{2} \left(m_e^2-m_\mu^2\right)^2 \pi ^2}C_{eZ}^{e\mu} +\left(e m_Z \left(m_e^4 \left(8-16 s_W^2\right)+3 m_\mu^2 m_Z^2 \left(1-4 s_W^2\right)\right. \right. \right.\nonumber \\
&+\left. \left. \left. 2 m_e^2 \left(m_\mu^2 \left(-1+4 s_W^2\right)+m_Z^2 \left(-3+8 s_W^2\right)\right)\right)\right)\frac{C_{eZ}^{\mu e} }{\left(32 \sqrt{2} \left(m_e^2-m_\mu^2\right)^2 \pi ^2\right)}
\right) \nonumber \\
&+B_0\left[m_e^2,m_\mu^2,m_H^2\right]\left(\frac{m_\mu^2 \left(m_e^2+m_H^2-3 m_\mu^2\right) m_W s_W}{32 \sqrt{2} \left(m_e^2-m_\mu^2\right)^2 \pi ^2}C_{e\varphi}^{\mu e}\right. \nonumber \\
&+\left.\frac{m_\mu \left(-2 m_e^2 m_H^2+\left(3 m_e^2+m_H^2\right) m_\mu^2-m_\mu^4\right) m_W s_W}{32 \sqrt{2} m_e \left(m_e^2-m_\mu^2\right)^2 \pi ^2}C_{e\varphi}^{e\mu} \right) \nonumber \\
&+B_0\left[m_e^2,m_\mu^2,m_Z^2\right]
\left(
\frac{e m_\mu m_Z^2 \left(3 m_e^2-3 m_\mu^2+2 \left(-3 m_e^2+m_\mu^2+m_Z^2\right) s_W^2\right) }{32 \left(m_e^2-m_\mu^2\right)^2 \pi ^2}C_{\varphi e} \right.\nonumber \\
&+
\left(e \left(m_\mu^2 \left(m_\mu^2-m_Z^2\right) \left(m_\mu^2+2 m_Z^2 \left(1-2 s_W^2\right)\right)+m_e^4 \left(3 m_\mu^2+4 m_Z^2 \left(-1+2 s_W^2\right)\right)\right. \right. \nonumber \\
&- \left. \left. m_e^2 \left(4 m_\mu^4+m_\mu^2 m_Z^2 \left(1-4 s_W^2\right)+4 m_Z^4 \left(-1+2 s_W^2\right)\right)\right)\frac{C_{\varphi l}^{(1)}+C_{\varphi l}^{(3)}}{64 m_e \left(m_e^2-m_\mu^2\right)^2 \pi ^2}\right. \nonumber\\
&+\left. \frac{e m_\mu m_Z \left(m_\mu^4-m_\mu^2 m_Z^2+m_e^2 \left(-3 m_\mu^2+2 m_Z^2\right)\right) \left(-3+4 s_W^2\right)}{32 \sqrt{2} m_e \left(m_e^2-m_\mu^2\right)^2 \pi ^2}C_{eZ}^{e\mu}\right. \nonumber \\
&+\left. \left(e m_Z \left(m_e^4 \left(4-8 s_W^2\right)+m_\mu^2 \left(3 m_Z^2 \left(1-4 s_W^2\right)+m_\mu^2 \left(3+4 s_W^2\right)\right)\right. \right. \right. \nonumber \\
&+\left. \left. \left. m_e^2 \left(4 m_Z^2 \left(-1+2 s_W^2\right)+m_\mu^2 \left(-5+12 s_W^2\right)\right)\right)\right)\frac{C_{eZ}^{\mu e}}{\left(32 \sqrt{2} \left(m_e^2-m_\mu^2\right)^2 \pi ^2\right)}
\right) \nonumber\\
&+B_0\left[m_\mu^2,0,m_W^2\right]\left(\frac{e m_e \left(2 m_W^2 \left(-2 m_\mu^2-m_W^2\right)+m_e^2 \left(m_\mu^2+3 m_W^2\right)\right)}{16 \left(m_e^2-m_\mu^2\right)^2 \pi ^2}C_{\varphi l}^{(3)} \right.\nonumber \\
&+\left.\frac{c_W^2 e m_e \left(2 m_\mu^2 m_W^2+m_e^2 \left(m_\mu^2-m_W^2\right)\right) m_Z}{16 \sqrt{2} m_\mu \left(m_e^2-m_\mu^2\right)^2 \pi ^2}C_{eZ}^{e\mu}
-\left(e \left(m_e^4 \left(m_\mu^2+3 m_W^2\right)\right. \right. \right.\nonumber \\
&+\left. \left. \left. m_\mu^2 \left(m_\mu^4-m_\mu^2 m_W^2+2 c_W^4 m_Z^4\right)-m_e^2 \left(2 m_\mu^4+3 m_\mu^2 m_W^2+3 c_W^4 m_Z^4\right)\right)\right)\frac{C_{eZ}^{\mu e}}{\left(16 \sqrt{2} \left(m_e^2-m_\mu^2\right)^2 m_Z \pi ^2\right)}
\right)  \nonumber \\
&+B_0\left[m_\mu^2,m_e^2,m_H^2\right]\left( \frac{m_e^2 \left(-3 m_e^2+m_H^2+m_\mu^2\right) m_W s_W}{32 \sqrt{2} \left(m_e^2-m_\mu^2\right)^2 \pi ^2}C_{e\varphi}^{\mu e}\right.\nonumber \\
&+\left.\frac{m_e \left(-m_e^4-2 m_H^2 m_\mu^2+m_e^2 \left(m_H^2+3 m_\mu^2\right)\right) m_W s_W}{32 \sqrt{2} m_\mu \left(m_e^2-m_\mu^2\right)^2 \pi ^2}C_{e\varphi}^{e\mu} \right) \nonumber \\
&+B_0\left[m_\mu^2,m_e^2,m_Z^2\right]\left(
\frac{e m_Z^2 \left(2 m_e^3-m_e m_Z^2+2 m_e \left(m_e^2-3 m_\mu^2+m_Z^2\right) s_W^2\right)}{32 \left(m_e^2-m_\mu^2\right)^2 \pi ^2}(C_{\varphi l}^{(1)}+C_{\varphi l}^{(3)})\right.\nonumber \\
&-\left.\frac{e m_e^2 (m_e-m_\mu) (m_e+m_\mu) \left(m_e^2-3 m_\mu^2-m_Z^2\right)+4 e \left(m_e^2-2 m_\mu^2\right) m_Z^2 \left(m_e^2+m_\mu^2-m_Z^2\right) s_W^2}{64 m_\mu \left(m_e^2-m_\mu^2\right)^2 \pi ^2}C_{\varphi e} \right.\nonumber\\
&+\left.\frac{e m_e m_Z \left(m_e^4+2 m_\mu^2 m_Z^2-m_e^2 \left(3 m_\mu^2+m_Z^2\right)\right) \left(1+4 s_W^2\right)}{32 \sqrt{2} m_\mu \left(m_e^2-m_\mu^2\right)^2 \pi ^2}C_{eZ}^{e\mu} +\left(e m_Z \left(-m_e^2 \left(5 m_e^2+m_\mu^2-3 m_Z^2\right)\right.\right.\right.\nonumber \\
&+\left.\left.\left.4 \left(m_e^4+3 m_e^2 (m_\mu-m_Z) (m_\mu+m_Z)-2 m_\mu^2 (m_\mu-m_Z) (m_\mu+m_Z)\right) s_W^2\right)\right)\frac{C_{eZ}^{\mu e}}{\left(32 \sqrt{2} \left(m_e^2-m_\mu^2\right)^2 \pi ^2\right)}
\right) \nonumber \\
&+B_0\left[m_\mu^2,m_\mu^2,m_H^2\right]\left( \frac{m_e m_\mu \left(m_H^2-2 m_\mu^2\right) m_W s_W}{32 \sqrt{2} \left(m_e^2-m_\mu^2\right)^2 \pi ^2}C_{e\varphi}^{e\mu}+\frac{\left(m_e^2-2 m_\mu^2\right) \left(m_H^2-2 m_\mu^2\right) m_W s_W}{32 \sqrt{2} \left(m_e^2-m_\mu^2\right)^2 \pi ^2}C_{e\varphi}^{\mu e} \right) \nonumber \\
&+B_0\left[m_\mu^2,m_\mu^2,m_Z^2\right]\left(
\frac{e m_Z^2 \left(3 m_\mu^4-4 m_\mu^2 m_Z^2 s_W^2+m_e^2 \left(2 m_Z^2 s_W^2+m_\mu^2 \left(-3+4 s_W^2\right)\right)\right)}{32 m_\mu \left(m_e^2-m_\mu^2\right)^2 \pi ^2}C_{\varphi e} \right.\nonumber \\
&+\left. \frac{e m_e \left(2 m_\mu^4+m_\mu^2 m_Z^2+2 m_Z^4 \left(-1+2 s_W^2\right)+m_e^2 \left(-2 m_\mu^2+m_Z^2 \left(3-8 s_W^2\right)\right)\right)}{64 \left(m_e^2-m_\mu^2\right)^2 \pi ^2}(C_{\varphi l}^{(1)}+C_{\varphi l}^{(3)}) \right.\nonumber \\
&+\left. \frac{e m_e m_\mu m_Z \left(2 m_\mu^2-m_Z^2\right) \left(-3+4 s_W^2\right)}{32 \sqrt{2} \left(m_e^2-m_\mu^2\right)^2 \pi ^2}C_{eZ}^{e\mu}+\left( e m_Z \left(m_e^2 \left(2 m_\mu^2-3 m_Z^2\right) \left(-1+4 s_W^2\right)\right.\right.\right.\nonumber \\
&-\left.\left.\left.2 m_\mu^2 \left(m_Z^2+8 (m_\mu-m_Z) (m_\mu+m_Z) s_W^2\right)\right)\right)\frac{C_{eZ}^{\mu e}}{\left(32 \sqrt{2} \left(m_e^2-m_\mu^2\right)^2 \pi ^2\right)}
\right),
\end{align}

\begin{align}
&\quad C_{TL}^{(C_0)}\nonumber \\
&=C_0\left[m_\mu^2,m_e^2,0,m_e^2,m_H^2,m_e^2\right]\left( -\frac{ m_e^3 m_\mu m_W s_W}{16 \sqrt{2} \left(m_e^2-m_\mu^2\right) \pi ^2}C_{e\varphi}^{e\mu}+\frac{ m_e^2 \left(2 m_e^2-m_\mu^2\right) m_W s_W}{16 \sqrt{2} \left(m_e^2-m_\mu^2\right) \pi ^2}C_{e\varphi}^{\mu e}\right) \nonumber \\
&+C_0\left[m_\mu^2,m_e^2,0,m_e^2,m_Z^2,m_e^2\right]\left(
-\frac{e m_e^2 m_\mu \left(m_e^2-m_\mu^2+4 m_Z^2 s_W^2\right)}{32 \left(m_e^2-m_\mu^2\right) \pi ^2}C_{\varphi e} \right.\nonumber \\
&+\left.\frac{e m_e^3 m_Z^2 \left(-1+2 s_W^2\right)}{16 \left(m_e^2-m_\mu^2\right) \pi ^2}(C_{\varphi l}^{(1)}+C_{\varphi l}^{(3)}) \right.\nonumber \\
&+\left.\frac{e m_e^3 m_\mu m_Z \left(1+4 s_W^2\right)}{16 \sqrt{2} \left(m_e^2-m_\mu^2\right) \pi ^2}C_{eZ}^{e\mu}-\frac{ e m_e^2 m_Z \left(m_\mu^2 \left(1-4 s_W^2\right)+m_e^2 \left(-4+8 s_W^2\right)\right)}{16 \sqrt{2} \left(m_e^2-m_\mu^2\right) \pi ^2}C_{eZ}^{\mu e}
\right) \nonumber \\
&+C_0\left[m_\mu^2,m_e^2,0,m_\mu^2,m_H^2,m_\mu^2\right]\left( \frac{m_e m_\mu^3 m_W s_W}{16 \sqrt{2} \left(m_e^2-m_\mu^2\right) \pi ^2}C_{e\varphi}^{e\mu}+\frac{ m_\mu^2 \left(m_e^2-2 m_\mu^2\right) m_W s_W}{16 \sqrt{2} \left(m_e^2-m_\mu^2\right) \pi ^2}C_{e\varphi}^{\mu e}\right) \nonumber \\
&+C_0\left[m_\mu^2,m_e^2,0,m_\mu^2,m_Z^2,m_\mu^2\right]\left(
\frac{e m_\mu^3 m_Z^2 s_W^2}{8 \left(-m_e^2+m_\mu^2\right) \pi ^2}C_{\varphi e}  \right.\nonumber \\
&+\left. \frac{e m_e m_\mu^2 \left(-m_e^2+m_\mu^2+2 m_Z^2 \left(1-2 s_W^2\right)\right)}{32 \left(-m_e^2+m_\mu^2\right) \pi ^2}(C_{\varphi l}^{(1)}+C_{\varphi l}^{(3)}) \right.\nonumber\\
&-\left.\frac{e m_\mu^2 m_Z \left(3 m_e m_\mu-4 m_e m_\mu s_W^2\right)}{16 \sqrt{2} \left(-m_e^2+m_\mu^2\right) \pi ^2}C_{eZ}^{e\mu}-\frac{e m_\mu^2 m_Z \left(m_e^2-4 \left(m_e^2-2 m_\mu^2\right) s_W^2\right)}{16 \sqrt{2} \left(-m_e^2+m_\mu^2\right) \pi ^2}C_{eZ}^{\mu e} 
\right) \nonumber \\
&+C_0\left[m_\mu^2,m_e^2,0,m_W^2,0,m_W^2\right]
\left(-\frac{e m_e m_W^2 \left(2 m_e^2-m_\mu^2-2 m_W^2\right) }{8 \left(m_e^2-m_\mu^2\right) \pi ^2}C_{\varphi l}^{(3)}  \right.\nonumber \\
&-\left.\frac{c_W^4 e m_e m_\mu m_Z^3}{8 \sqrt{2} \left(m_e^2-m_\mu^2\right) \pi ^2}C_{eZ}^{e\mu} 
+\frac{c_W^2 e \left(2 m_e^4+m_\mu^2 m_W^2-2 m_e^2 \left(m_\mu^2+m_W^2\right)\right) m_Z}{8 \sqrt{2} \left(m_e^2-m_\mu^2\right) \pi ^2}C_{eZ}^{\mu e} 
\right),
\end{align}

\begin{align}
&\quad C_{TL}^{(c)}\nonumber \\
&=
\frac{e m_\mu \left(-m_e^2+m_\mu^2+8 m_Z^2 s_W^2\right)}{64 \left(-m_e^2+m_\mu^2\right) \pi ^2}C_{\varphi e}+
\frac{e m_e \left(-m_e^2+m_\mu^2+4 m_Z^2 \left(-1+2 s_W^2\right)\right)}{64 \left(m_e^2-m_\mu^2\right) \pi ^2}C_{\varphi l}^{(1)} \nonumber \\
&-\frac{e m_e \left(m_e^2+3 m_\mu^2-4 m_Z^2\right)}{64 \left(m_e^2-m_\mu^2\right) \pi ^2}C_{\varphi l}^{(3)}+\frac{m_W s_W}{32 \sqrt{2} \pi ^2}C_{e\varphi}^{\mu e}-\frac{\sqrt{2}  m_W s_W}{e}C_{e\gamma}^{\mu e} \nonumber \\
&+\frac{e m_e m_\mu m_Z \left(1+s_W^2\right)}{16 \sqrt{2} \left(m_e^2-m_\mu^2\right) \pi ^2}C_{eZ}^{e\mu}+\frac{e m_Z \left(m_\mu^2 \left(7-2 s_W^2\right)+m_e^2 \left(-5+4 s_W^2\right)\right)}{32 \sqrt{2} \left(m_e^2-m_\mu^2\right) \pi ^2}C_{eZ}^{\mu e} \nonumber \\
&+\frac{e}{16 \pi ^2}\left(m_e C_{le}^{e\mu ee}+m_\mu C_{le}^{\mu \mu e\mu}+ m_\tau C_{le}^{\mu \tau\tau e}\right),
\label{CTL0}
\end{align}

Note that \Eqn{swap} applied to \Eqn{CTL0} also implies that the
generation indices in the operators $C_{le}$ have to be swapped.

\section{Lepton-flavour violating $\tau$ decays and effective coefficient constraints} 
\label{App:c}
\setcounter{equation}{0} 
\noindent
In this appendix, the strategy adopted in the main text is extended to
the case of lepton-flavour violating tauonic transitions. By combining
(see \cite{Beringer:1900zz}) the experimental values obtained at the
LEP collider (see \cite{Alexander:1996nc, Balest:1996cr,
Barate:1997be, Acciarri:2000vq, Abdallah:2003yq}), the $\tau$-lepton
total width is inferred to be
\begin{align}\label{gammatau}
\Gamma_\tau=2.3\cdot 10^{-12}\ \GeV.
\end{align}

Recently, the BaBar Collaboration established \cite{Aubert:2009ag} the
following limits on the tauonic lepton-flavour violating decay
rates\footnote{Somewhat weaker limits have been obtained by the Belle
  collaboration~\cite{Hayasaka:2007vc}.}:
\begin{align}\label{teg:lim}
{\rm BR}(\tau^-\to e^-\gamma)&\leq 3.3\cdot 10^{-8},\\
\label{tmg:lim}
{\rm BR}(\tau^-\to \mu^-\gamma)&\leq 4.4\cdot 10^{-8}.
\end{align}

Putting together the information in \Eqns{gammatau}{tmg:lim} and
adapting \Eqn{BRanalitica} of Section~\ref{Sec:3} to the tauonic case,
the following limits are obtained: 
\begin{align}
\fbox{$\tau\to e\gamma$} & \Longrightarrow 
\left. \frac{\sqrt{|C_{TL}(\lambda)|^2+|C_{TR}(\lambda)|^2}}{\Lambda^2}
\right|_{\lambda\ll \Lambda}\leq 4.1\cdot 10^{-10}\left[\GeV\right]^{-1},\\[2ex]
\fbox{$\tau\to \mu\gamma$} & \Longrightarrow 
\left. \frac{\sqrt{|C_{TL}(\lambda)|^2+|C_{TR}(\lambda)|^2}}{\Lambda^2}
\right|_{\lambda\ll\Lambda}\leq 4.7\cdot 10^{-10}\left[\GeV\right]^{-1}.
\end{align}

The functional form of the coefficients $C_{TL}$ and $C_{TL}$ is not
different from the result of Table~\ref{tab:res}, apart from suitable
changes of the mass parameters and generation indices (e.g. for the
$\tau\to e\gamma$ case one should replace $m_\mu$ with $m_\tau$ except
for the contribution from $Q_{le}$). Hence, exploiting the strategy
that was presented in Section~\ref{Sec:4}, a set of both fixed-scale
and $\Lambda$-dependent limits can be obtained for new coefficients
involving a LFV connected to the third generation.  Similarly to what
has been done already, such results are summarised in
Tables~\ref{tab:fo_teres}-\ref{tab:rge_tmres}. A final remark is
required: as in \Eqn{CeyatmZ} the limits on $C_{e\gamma}$ at the $m_Z$
scale are slightly different from the ones at the $m_\tau$ scale
presented in Tables~\ref{tab:fo_teres}~and~\ref{tab:fo_tmres}. In
fact, the limits evaluated at the electroweak scale read
\begin{align}
\sqrt{\frac{
|C_{e\gamma}^{\tau e}(m_Z)|^2 +|C_{e\gamma}^{e\tau}(m_Z)|^2}{2}}
\leq 1.7\cdot 10^{-12}\frac{\Lambda^2}{\left[\GeV\right]^2}
\label{Ctaue},\\[2ex]
\sqrt{\frac{
|C_{e\gamma}^{\tau \mu}(m_Z)|^2 +|C_{e\gamma}^{\mu\tau}(m_Z)|^2}{2}}
\leq 2.0\cdot 10^{-12}\frac{\Lambda^2}{\left[\GeV\right]^2}.
\label{Ctaumu}
\end{align}
Applying the RG evolution and using \Eqns{Ctaue}{Ctaumu}, one can
extract the values of
Tables~\ref{tab:rge_teres}~and~\ref{tab:rge_tmres}.

\begin{table}[!ht] 
\centering
\renewcommand{\arraystretch}{1.2}
\btb{||c|c||c|c||} 
\hline 
\hline
\multicolumn{4}{||c||}{$\boldsymbol{\tau\to e\gamma}$}\\ 
\hline
3-P Coefficient & At fixed scale & 4-P Coefficient & At fixed scale  \\
\hline 
$C_{e\gamma}^{\tau e}$ & $2.4\cdot
10^{-12}\frac{\Lambda^2}{\left[\GeV\right]^2}$ & 
$C_{le}^{\tau eee}$ & $4.2\cdot 10^{-4}\frac{\Lambda^2}{\left[\GeV\right]^2}$ \\
$C_{eZ}^{\tau e}(m_Z)$ &
$1.3\cdot10^{-9}\frac{\Lambda^2}{\left[\GeV\right]^2}$ & 
$C_{le}^{ \tau\mu\mu e}$ & $2.0\cdot 10^{-6}\frac{\Lambda^2}{\left[\GeV\right]^2}$ \\
$C_{\vp l}^{(1)}$ &
$1.5\cdot10^{-7}\frac{\Lambda^2}{\left[\GeV\right]^2}$ & 
$C_{le}^{\tau \tau\tau e}$ & $1.2\cdot 10^{-7}\frac{\Lambda^2}{\left[\GeV\right]^2}$ \\
$C_{\vp l}^{(3)}$ &
$1.4\cdot10^{-7}\frac{\Lambda^2}{\left[\GeV\right]^2}$ &
 &  \\
$C_{\vp e}$ &
$1.4\cdot10^{-7}\frac{\Lambda^2}{\left[\GeV\right]^2}$ &
 & \\
$C_{e\vp}^{\tau e}$ &
$1.7\cdot10^{-6}\frac{\Lambda^2}{\left[\GeV\right]^2}$ & &
\\[1ex]
\hline 
\hline
\etb
\caption{Limits on the Wilson coefficients contributing to the
 $\tau\to e\gamma$ transition up to the one-loop
 level. \label{tab:fo_teres}} 
\end{table}

\begin{table}[!ht] 
\centering
\renewcommand{\arraystretch}{1.2}
\btb{||c||c|c|c||} 
\hline 
\hline
\multicolumn{4}{||c||}{$\boldsymbol{\tau\to e\gamma}$}\\ 
\hline 
3-P Coefficient & at  $\Lambda = 10^3~\GeV$ &
at  $\Lambda = 10^4~\GeV$ & at  $\Lambda = 10^5~\GeV$ \\
\hline 
$C_{e\gamma}^{\tau e}$ & $2.5\cdot10^{-6}$ & $2.6\cdot10^{-4}$ & $2.8\cdot10^{-2}$ \\
$C_{eZ}^{\tau e}$ & $2.3\cdot10^{-4}$   & $1.3\cdot10^{-2}$ & $9.5\cdot10^{-1}$ \\
$C^{(3)}_{\tau ett}$ &  $3.4\cdot10^{-5}$ & $1.9\cdot10^{-3}$  & $1.4\cdot10^{-1}$  \\
$C^{(1)}_{\tau ett}$ &  $1.8\cdot10^{-2}$ & $5.0\cdot10^{-1}$  & n/a \\
$C^{(3)}_{\tau ecc}$ &  $4.6\cdot10^{-3}$ &  $2.5\cdot10^{-1}$ & n/a  \\
$C^{(1)}_{\tau ecc}$ & $\sim 2.4$  &  n/a  &  n/a \\[1ex]
 \hline 
\hline
\etb
\caption{Limits on the Wilson coefficients defined at the scale
$\lambda=\Lambda$ for three choices of $\Lambda =10^3, 10^4,
10^5~\GeV$.  \label{tab:rge_teres}}
\end{table}


\begin{table}[!ht] 
\centering
\renewcommand{\arraystretch}{1.2}
\btb{||c|c||c|c||} 
\hline 
\hline
\multicolumn{4}{||c||}{$\boldsymbol{\tau\to \mu\gamma}$}\\ 
\hline
3-P Coefficient & At fixed scale  & 4-P Coefficient & At fixed scale  \\
\hline 
$C_{e\gamma}^{\tau\mu}$ & $2.7\cdot
10^{-12}\frac{\Lambda^2}{\left[\GeV\right]^2}$ &  $C_{le}^{\tau ee\mu}$ & 
 $4.8\cdot 10^{-4}\frac{\Lambda^2}{\left[\GeV\right]^2}$ \\
$C_{eZ}^{\tau\mu}(m_Z)$ &
$1.5\cdot10^{-9}\frac{\Lambda^2}{\left[\GeV\right]^2}$ &
$C_{le}^{ \tau\mu \mu\mu}$ &  $2.3\cdot 10^{-6}\frac{\Lambda^2}{\left[\GeV\right]^2}$ \\
$C_{\vp l}^{(1)}$ &
$1.7\cdot10^{-7}\frac{\Lambda^2}{\left[\GeV\right]^2}$ &
$C_{le}^{\tau \tau\tau \mu}$ &  $1.4\cdot 10^{-7}\frac{\Lambda^2}{\left[\GeV\right]^2}$ \\
$C_{\vp l}^{(3)}$ &
$1.6\cdot10^{-7}\frac{\Lambda^2}{\left[\GeV\right]^2}$ & & \\
$C_{\vp e}$ &
$1.6\cdot10^{-7}\frac{\Lambda^2}{\left[\GeV\right]^2}$ & & \\
$C_{e\vp}^{\tau\mu}$ &
$1.9\cdot10^{-6}\frac{\Lambda^2}{\left[\GeV\right]^2}$ & & 
 \\[1ex]
\hline 
\hline
\etb
\caption{Limits on the Wilson coefficients contributing to the
 $\tau\to \mu\gamma$ transition up to the one-loop
 level.  \label{tab:fo_tmres}}
\end{table}

\begin{table}[!ht] 
\centering
\renewcommand{\arraystretch}{1.2}
\btb{||c||c|c|c||} 
\hline 
\hline
\multicolumn{4}{||c||}{$\boldsymbol{\tau\to \mu\gamma}$}\\ 
\hline 
3-P Coefficient & at  $\Lambda = 10^3~\GeV$ &
at  $\Lambda = 10^4~\GeV$ & at  $\Lambda = 10^5~\GeV$ \\
\hline 
$C_{e\gamma}^{\tau\mu}$ & $3.0\cdot10^{-6}$ & $3.1\cdot10^{-4}$ & $3.2\cdot10^{-2}$ \\
$C_{eZ}^{\tau\mu}$ &  $2.8\cdot10^{-4}$   & $1.5\cdot10^{-2}$ & $\sim 1.1$ \\
$C^{(3)}_{\tau\mu tt}$ & $4.0\cdot10^{-5}$ & $2.2\cdot10^{-3}$  & $1.6\cdot10^{-1}$  \\
$C^{(1)}_{\tau\mu tt}$ &  $2.1\cdot10^{-2}$ &  $5.9\cdot10^{-1}$ &  n/a \\
$C^{(3)}_{\tau\mu cc}$ &  $5.4\cdot10^{-3}$ &  $3.0\cdot10^{-1}$ & n/a  \\
$C^{(1)}_{\tau\mu cc}$ & $\sim 2.8$  &  n/a  &  n/a \\[1ex]
 \hline 
\hline
\etb
\caption{Limits on the Wilson coefficients defined at the scale
  $\lambda=\Lambda$ for three choices of $\Lambda =10^3, 10^4,
  10^5~\GeV$. \label{tab:rge_tmres}}
\end{table}

\clearpage

\bibliographystyle{apsrev}
\bibliography{biblio}

\end{document}